\documentclass[a4paper,fleqn]{cas-dc}

\usepackage[numbers]{natbib}
\usepackage{amssymb}
\usepackage{amsmath}
\usepackage{bm}
\usepackage{tikz}
\usepackage{amssymb,amsmath,amsthm}
\usetikzlibrary{shapes,arrows,positioning,calc}
\usepackage{makecell}
\usepackage{pifont}
\usepackage{color, colortbl}
\usepackage{multirow}
\usepackage[ruled]{algorithm2e}
\usepackage{arydshln} % for the hdashline
\usepackage{bbm} % for \mathbbm{1}
\usepackage[caption=false,font=normalsize,labelfont=sf,textfont=sf]{subfig}
\usepackage{enumitem}

\newcommand{\vect}[1]{\mathbf{#1}}

\definecolor{lightgray}{rgb}{0.83, 0.83, 0.83}
\definecolor{darkgreencomment}{RGB}{0, 153, 76}
\definecolor{burntorange}{rgb}{0.8, 0.33, 0.0}
\definecolor{slateblue}{RGB}{106, 90, 205}

\newtheorem{problem}{Problem}
\newtheorem{definition}{Definition}
\newtheorem{example}{Example}
\newtheorem{rem}{Remark}
\newtheorem{prop}{Proposition}
\newtheorem{corol}{Corollary}
\newtheorem{assumption}{Assumption}
\DeclareMathOperator{\Ima}{Im}
\DeclareMathOperator*{\argmin}{arg\,min}
\DeclareMathOperator*{\argmax}{arg\,max}
\renewcommand{\mathbbm}[1]{\text{\usefont{U}{bbm}{m}{n}#1}}

\usepackage[normalem]{ulem}
\usepackage{soul}

\usepackage[color=yellow, textsize=small, textwidth=\marginparwidth, draft]{todonotes}   % notes 

\begin{document}
\let\WriteBookmarks\relax
\def\floatpagepagefraction{1}
\def\textpagefraction{.001}
\shorttitle{Visual Sensor Network Stimulation Model Identification via Gaussian Mixture Model and Deep Embedded Features}    
\shortauthors{L. Varotto et al.}  
\title [mode = title]{Visual Sensor Network Stimulation Model Identification\\ via Gaussian Mixture Model and Deep Embedded Features}  

% Titolo originale:
% Visual Sensor Network SM Identification via Deep Embedded Features and Gaussian Mixture Models

% Title footnote mark
% eg: \tnotemark[1]
%\tnotemark[1] 
% Title footnote 1.
% eg: \tnotetext[1]{Title footnote text}
%\tnotetext[1]{Title footnote text}

% First author

\author[1]{Luca Varotto}[orcid=0000-0002-2450-9106]
\cormark[1]
\ead{luca.varotto.5@phd.unipd.it}
\affiliation[1]{organization={Department of Information Engineering, University of Padova},
	addressline={via Gradenigo 6/B}, 
	city={Padua},
	postcode={35131}, 
	country={Italy}}

\author[1]{Marco Fabris}[orcid=0000-0003-1728-1331]
\ead{marco.fabris.7@phd.unipd.it}
\author[2]{Giulia Michieletto}[orcid=0000-0002-1357-8077]
\ead{giulia.michieletto@unipd.it}
\author[1]{Angelo Cenedese}[orcid=0000-0003-2249-5094]
\ead{angelo.cenedese@unipd.it}
\affiliation[2]{organization={Department of Management and Engineering, University of Padova},
            addressline={Stradella San Nicola, 3}, 
            city={Vicenza},
            postcode={36100}, 
            country={Italy}}

\begin{abstract}
Visual sensor networks (VSNs) constitute a fundamental class of distributed sensing systems, with unique complexity and appealing performance features, which correspondingly bring in quite active lines of research. %subjects. 
An important research direction consists in the identification and estimation of the VSN sensing features: 
%of a mathematical representation of the VSN
these are 
%informative and 
practically useful 
%in particular 
when scaling with the number of 
%VSN 
cameras or with the observed scene complexity. 
%One of these novel challenges is represented by 
%This paper tackles the identification of the network stimulation model (SM), which emerges when a set of detectable events trigger different subsets of the cameras. In this direction, the formulation of the related SM identification problem is proposed, along with a proper network observations generation method.
%
With this context in mind, this paper introduces for the first time the idea of Stimulation Model (SM), 
%is introduced in this paper, which emerges when 
as a mathematical relation between the set of detectable events and the corresponding stimulated cameras observing those events. 
%the VSN as an ensemble and different and disjoint subsets of cameras need to be associated to the single event according to the quality of their observations.
%
The formulation of the related SM identification problem is proposed, along with a proper network observations model, and a solution approach based on deep embedded features and soft clustering. 
%is leveraged. %to solve the presented identification problem. 
In detail: first, the Gaussian Mixture Modeling is employed to provide a suitable description for data distribution, while an autoencoder is used to reduce undesired effects due to the so-called curse of dimensionality emerging in case of large scale networks. 
%Hence, 
Then, it is shown that a SM can be learnt by solving Maximum A-Posteriori estimation on the encoded features belonging to a space with lower dimensionality. %\ACe{thus providing a descriptive model for the number of observed active events together with a occurrence probability}. 
%
%Lastly, 
Numerical results on synthetic scenarios are reported to validate the devised estimation algorithm. 
\end{abstract}

% Use if graphical abstract is present
%\begin{graphicalabstract}
%\includegraphics{}
%\end{graphicalabstract}

% Research highlights
%\begin{highlights}
%\item 
%\item 
%\item 
%\end{highlights}

% Each keyword is seperated by \sep

%\journal{Engineering Applications of Artificial Intelligence}
%\hyphenation{op-tical net-works semi-conduc-tor}
\maketitle

% INTRODUCTION
\section{Introduction}\label{sec:intro}

% INTRO ON VSNs
%A Visual Sensor Network (VSN) is a multi-sensor system composed of a collection of spatially distributed camera-devices capable of communicating over a wireless network \cite{kyung2016theory,aghajan2009multi,soro2009survey}. 
{The recent boost towards the development of Internet-of-Things architectures is leading to the substitution of the traditional monitoring and surveillance camera networks in favor of smart Visual Sensor Networks (VSNs). From an overall perspective, such 
%\sout{highly cooperative} 
multi-sensor systems are composed of a collection of spatially distributed static or (partially) dynamic smart camera-devices, capable of communicating over a wireless network both raw images and complex, semantic data, obtained through advanced computations. Thus, VSNs are, potentially, highly cooperative systems where images are fused  
%\sout{and then processing and fusing images of a scene} 
from a variety of viewpoints, whenever it turns out to be %\sout{into some form} 
more useful than extracting information from individual images~\cite{kyung2016theory}. %, \cite{aghajan2009multi,soro2009survey}.%,aghajan2009multi,soro2009survey
} 
% Thanks to the advances in high performance embedded microcontrollers \cite{pohl2016advanced}, optimized computer vision techniques \cite{voulodimos2018deep} and reliable communication protocols \cite{li2019survey}, multi-camera networks have increasingly spread out in the last twenty years becoming ubiquitous smart systems in industrial, civil and domestic contexts \cite{aghajan2009multi}.  
% By acquiring information-rich data, these architectures enable vision-based interpretative applications as intelligent IoT surveillance \cite{tahir2014cost}, smart living domotics \cite{fleck2008smart}, autonomous vehicles and assisted driving \cite{janai2017computer}, industrial environments monitoring and control \cite{LuvisottoClustering}, urban perimeter/area patrolling for events detection \cite{bof2017asynchronous}, to cite a few.
{Supported by the last advances in high performance embedded microprocessors, fully optimized computer vision algorithms and safe, fast and reliable communication protocols, VSNs are rapidly becoming an ubiquitous and strategic technology in industrial, rural, civil and domestic contexts, enabling vision-based interpretative applications as, for instance, events and crowd monitoring~\cite{singh2021crowd}, %, \cite{bisagno2018dynamic}, %bisagno2018dynamic 
dynamic intruders tracking~\cite{lissandrini2019cooperative}, and assisted and autonomous driving~\cite{liu2017robust}.}

% DEFINITION AND MOTIVATION TOWARDS NETWORK TOPOLOGY (VISION AND TRANSITION MODELS - FOCUS ON TRANSITION MODELS) LEARNING
% In all the mentioned scenarios, the efficiency related to the task completion is strictly conditioned by the knowledge  \LV{(completely copied from \cite{VarottoECC19})} of the network coverage model \cite{mavrinac2013modeling}, \ToDo{which represents the ... .}
%\sout{All the mentioned application scenarios benefit from the knowledge of the coverage properties of the employed VSN, but also of the field of view (FoV) each sensor composing the network and of the relationship between the cameras in terms of their coverage capabilities.}
All the mentioned application scenarios benefit from the knowledge of each sensor's field of view (FoV) and, more in general, of the coverage capabilities of the entire VSN.
% \ToDo{In particular, it is called \textit{vision graph} when ... \cite{}, while it is referred to as \textit{transition graph} when ... \cite{}.}  
% \LV{\textit{Vision graph} is the \textit{sensing graph} (related to physical properties); \textit{semantic/stimulation graph} is as a \textit{transition graph} (i.e., related to the dynamics of observed events, not requiring overlap), when multiple semantic events are observed.}
%\sout{In particular,} 
When modeling the coverage features of a VSN, one can distinguish between the \textit{coverage overlap models} and the \textit{transition models}~\cite{mavrinac2013modeling}. 
The former focus on the physical relationships between pairs or groups of cameras having overlapping FoVs,~and can be obtained during a preliminary setup phase from the pose of the cameras in the environment and their viewing properties. The latter take into account the more abstract relationships arising from the 
%\sout{possibility or} 
probability of an event that is engaging in time multiple cameras FoVs~(e.g., the presence of a target moving across the environment). % a moving target passes from one camera fov to that of another~\cite{mavrinac2013modeling}. 
The coverage overlap models can be mathematically represented by undirected graphs wherein each node corresponds to a sensor in the network and the edges are defined according to the mentioned physical FoVs overlapping relations. These graphs are often referred to as \textit{visual sensing graphs}. Similarly, the transition models can be formalized through the \textit{transition graphs}, wherein the node set represents the devices group and the edge set is derived accounting for both the cameras geometric coverage and the environment dynamics model.

A particular type of transition graph is the \textit{stimulation graph}, wherein an edge between two nodes does not refer to a (non-zero) transition probability between the two corresponding cameras' FoVs, but it represents the fact that the same event can trigger (namely, stimulate) both the two cameras, either simultaneously or at different time instants. In particular, when multiple events can be observed by the same VSN, the stimulation graph results to be a multi-graph, namely a set of graphs, each of which is associated to a specific event.
% APPLICATIONS EXPLOITING NETWORK TOPOLOGY KNOWLEDGE
% The knowledge on (semantic) sensing topology is the basis for collaborative
% tasks like resource (e.g., energy, bandwidth) management \cite{dieber2011resource,aziz2013energy,li2013adaptive,esterle2014socio}, network self-organization (e.g., cluster head election)\cite{sanmiguel2014self}, distributed estimation strategies\cite{VarottoECC19,sorrentino2020group,tron2011distributed}, multi-camera tracking\cite{tesfaye2019multi,taj2011distributed}, and distributed and federated learning \cite{yang2019federated}.
%
%\sout{Also,} 
The stimulation graph is a graphical model representation of a 
%\sout{more general concept of} 
\textit{Stimulation Model} (SM),
%\sout{Indeed, the rationale behind the SM is that several events can be observed with different quality and capacity by different cameras: in order to improve the scene knowledge, there is the need}
which is 
%\sout{to learn } 
a map between the stimuli (events) and the information processing units (cameras). 
%\sout{ that provides an optimal association, which is exactly the SM}
%
The identification of such a 
%\sout{graph} 
map turns out to be advantageous when dealing with several collaborative VSN assignments, as, for instance, resources (e.g., energy, bandwidth) allocation and management~\cite{esterle2014socio}, \cite{dieber2011resource}, %,aziz2013energy,li2013adaptive}}, %dieber2011resource,aziz2013energy,li2013adaptive,
network self-organization requirements (e.g., cluster head election)~\cite{sanmiguel2014self}, distributed estimation~\cite{VarottoECC19}, \cite{sorrentino2020group}, %,,sorrentino2020group}, %,tron2011distributed
tracking tasks~\cite{tesfaye2019multi}, and distributed and federated learning actions~%\cite{ZhangXieBai2021}, 
\cite{yang2019federated}. %,yang2019federated 
Although its potentialities in the application context, the SM is a concept that has not been properly investigated in the existing VSN literature. 

The main contribution of this work, thus, consists in the formal definition of the \textit{SM identification problem}, and subsequently in the description of a proper solution. In detail, the outlined identification strategy takes advantages of the flexibility and efficient modeling capabilities characterizing the Gaussian Mixture Models (GMMs). These statistical tools, indeed, permit to approximate any given probability density with high accuracy~\cite{tzikas2008variational}; and, for this reason, they are exploited in a large variety of applications ranging from path planning~%\cite{ChengZhangLiu2021pathplanningGMM}, 
\cite{GMM_search_static,GMM_search_RHC},
to object tracking~\cite{GMM_tracking}, from image modeling and segmentation \cite{yan2021gmm}, \cite{GMM_BS}, %,GMM_image}} %GMM_BS,GMM_image,
to speech understanding~\cite{GMM_speech}.
Coping with the SM identification problem, a GMM is adopted for describing the VSN collected data, thus leading to the design of a learning based strategy, which involves the presence of an auto-encoder (AE) to deal with the emergence of possible dimensionality issues.

% RELATED WORKS
\subsection{Related Works}\label{subsec:related_works} 

% WORKS ON TOPOLOGY LEARNING

The literature devoted to the modeling of multi-camera systems is substantial~
{\cite{mavrinac2013modeling,zou2007determining,Hussain2021multiview}}, and  
%\sout{In this sense, a thorough survey of the state-of-the-art VSN  models is yielded in~\cite{mavrinac2013modeling}.} 
particularly in~\cite{mavrinac2013modeling}, the existing topological and geometric coverage models are investigated and compared considering the target applications. %, entailing the derivation of  an hypothetical inclusively general sensing model. 
Quite recently, also supported by the progression in the learning field, the identification of the VSNs topological features continues to represent a stimulating research topic (see, e.g.~\cite{SpagnoloAghajanBebis2021,WongCicekSoatto2021,liu2021multi}). %It is worth to notice that 
The interest is still mainly directed on the study of the coverage overlap models and the corresponding graph-based representations, which turns out to be a (complex but) static problem. 
The attention toward this issue  is principally justified by the determination of the optimal camera placement in order to face the traditional VSN surveillance and monitoring operations~\cite{altahir2017modeling,kritter2019optimal,han2019camera}. 
%
%\sout{On the other hand, however, the attention is recently focusing also toward the identification of more complex topological properties, deriving from the need of coping with highly dynamic events. For example, several camera network topology inference methods have been proposed for large-scale person re-identification}~\cite{cheng2019data,cho2019joint}.
%
On the other hand, however, the attention is also focused on the identification of topological properties  deriving from highly dynamic scenarios, outlining the importance of transition graphs in such contexts. 
For example, camera network topology inference methods have been proposed for large-scale person re-identification~\cite{cheng2019data,cho2019joint}, which infer the VSN topology based on the occurrence correlation between the people's entry and exit events.

%It is worth to note that,
%When accounting for
In the transition models definition, two elements result to be very important. On one side, communities and clusters arise as  important structural characteristics of the large VSNs, given that these consist in cameras subsets whose components are characterized by some common functional and organizational properties~\cite{LuvisottoClustering,javed2018community}. 
%
%\sout{A very close topic to transition models determination is the semantic identification of topologies, communities and layered networks, intended as the recognition of some common features among elements of complex systems. Also in this case, the statistical approach is generally envisaged. For instance, in~\cite{jin2020identification} a Bayesian model and an efficient variational inference algorithm are proposed to capture the generalized communities and the topical clusters in an independent fashion. 
%}
%
On the other side, the adopted network observations model is fundamental, since it defines the measurements characteristics, describes the data generation process, and accounts for the principal sources of data artifacts~\cite{milanfar2017super}. In this respect, probabilistic approaches are usually employed for modeling the observations in order to suitably represent realistic perception with uncertainties and distortions~\cite{varotto2021active}, and statistical filtering techniques are often used to retrieve the variables of interest~\cite{james2013introduction}.
Along this line, the approach proposed in~\cite{cenedese2010graph} for the identification of the transition graph describes the observable areas of a VSN relying on the hidden Markov model theory and,  similarly, in~\cite{lucchese2014hidden} the transitions %between any two states
are described by the distribution of an underlying Markov process; 
in~\cite{farrell2007learning}, instead, the transition model identification is faced by casting the network observations in a Bayesian framework.

In these works, however, the focus of the modeling is on the VSN components and the observations are exploited as a mean to establish a link among two or more cameras, which reveals to be fundamental in coordinated tasks of patrolling or tracking. 
Differently, in the SM proposed here, the observations take a role as important as those of the devices, and the aim of the model is to understand which sensor is more apt to observe which event.
Interestingly, this employment of the observations constitutes also a semantic representation of the VSN and its camera partitions: the association between cameras and events allow the specialization of any sensor and may yield performance improvements.

%\sout{%More in general, {In the recent years}, the problem of identifying a model of a sensor camera network from observations {has been often tackled by means of learning-based techniques.}
%considered in many research areas. 
%In \cite{cenedese2010graph}, {Authors in \cite{cenedese2010graph} exploit hidden Markov models to reconstruct}
%, it is reconstructed  the graph where the vertices represent states that are observable and distinguishable by the sensor network and edges are the feasible transitions among these states.  Moreover, in a follow up of this work, the problem of building a transitional model of an initially uncalibrated camera network \cite{lucchese2014hidden} is tackled through a Hidden-Markov-Model-based %technique, in which {method wherein} the model's state-space is characterized as a partition of the physical network coverage. 
%Transitions between any two such states are described by the distribution of the underlying Markov Process. 
%Authors in  In \cite{cheng2006determining}, it is presented a decentralized {approach leveraging principal component analysis} for retrieving the vision graph for a distributed, ad-hoc camera network, in which each edge of the graph models two cameras that monitor a sufficiently broad part of the same scene.}
%
Despite the diverse employment of the observations, how different approaches deal with the needed data volume for modeling is a further aspect discussed in the literature: indeed, the recent VSNs literature also accounts for the issues deriving from the increasing number of the devices composing the networks and the dataset dimensionality that follows.
%
%\sout{Several dimensionality reduction techniques are, indeed, well-known~\cite{dimensionalityReduction_tutorial, dimensionalityReduction_review}.  Among them, the solutions based on the AEs (see, e.g.~\cite{wang2016auto,DL_tutorial, charte2018practical}) are characterized by well-stated theoretical function approximation properties and demonstrated learning capabilities since these are based the deep neural networks~\cite{bengio2013representation}.}
%
In \cite{cheng2006determining}, e.g., it is presented a decentralized approach leveraging Principal Component Analysis for data compression to retrieve the vision graph of an ad-hoc camera network, in which each edge of the graph models two cameras that monitor a sufficiently broad part of the same scene.
More recently, \cite{huang2018incorporating} illustrates an effective method that encompasses latent structural constraints into binary compressed sensing, demonstrating high accuracy and robust effectiveness with 
%\sout{of such a technique by analyzing } 
artificial small-world and scale-free networks, as well as empirical networks. This strategy allows network construction based on evolutionary game data, it requires a relatively small number of observations and it is robust against strong measurement noise.
A general approach for dimensional reduction using iteratively thresholded ridge regression screeners has been developed in \cite{shi2020recovering}. After drastically diminishing the dimensions of the problem, the lasso approach is then used to recover the network connections.

%\sout{Finally, in \cite{spyrou2018weighted}, graph metrics such as degree, average neighbor degree, transitivity and modularity are used in network estimation by discovering optimization methods that encapsulate prior knowledge of the network structure. The derivatives of graph metrics are then employed in gradient descent algorithms for weighted undirected network decomposition, network denoising, network and completion.} 

The approach proposed in this paper for data dimensionality reduction differs from the solutions described so far, being based on the use of AEs~\cite{charte2018practical}.
As a matter of fact, AEs have been employed in VSN applications for specific tasks (e.g. for 3D motion understanding~\cite{lai2021video}, visual tracking~\cite{cheng2019visual}), but they have not been used (to the best of the authors knowledge) for the VSN model identification.
In this context, the AE approach appear quite promising, being characterized by well-stated theoretical function approximation properties and having demonstrated efficient and robust learning capabilities.

\subsection{Contributions}\label{subsec:contribs} 

Within the VSN context, this work focuses on the aforementioned SM, according to the cutting-edge interest toward the definition of transition models rather than coverage overlap models. To the best of authors knowledge, such a model has not been investigated in the existing literature, even though it turns out to be beneficial when dealing with complex highly dynamic scenarios. In particular, the novelty brought by this investigation is twofold:% and it is discussed in the following lines.}

(i) %The first contribution of this work %, thus, 
%consists in 
first, the rigorous definition of the Stimulation Model is provided. %Note that this concept has never been properly introduced in the existing VSN literature. 
Then, the resolution of the SM identification problem is formally stated. In detail, the latter consists in the estimation of a certain Boolean matrix, which encodes the stimulation pattern of the network components  by a set of events,  
%It is then shown that the SM identification problem %the has to be solved 
by exploiting the cameras collected data in correspondence to the occurrence of the events, namely by resting on the \textit{network observations}. 

(ii) %A second contribution is represented by the %A  subsequent 
the detailed characterization of both the events set and the network observation is derived in probabilistic terms leading to the design of a SM identification procedure 
based on some deep embedded features. More in detail, 
the outlined identification solution exploits the GMM interpretation of the given network observations model and envisages an AE to face the possible dimensionality issues, while solving a Maximum A-Posteriori estimate on the encoded features. 

The choice of exploiting GMMs is motivated by the fact that these are a valuable statistical tool for modeling densities. In particular, these are flexible enough to approximate any given density with high accuracy, hence they turn out to be a suitable approximate model for the cameras' observations.
%and, in particular, it will be shown that cameras' observations are, indeed, approximately distributed according to a GMM.
%
On the other side, AEs have been proven to be a valuable technique for dimensionality reduction, with superior performance with respect to both linear (e.g., Principal Component Analysis) and non-linear (e.g., isomap) methods~\cite{wang2016auto}.

The effectiveness of the outlined SM identification solution is assessed by evaluating some suitably defined performance metrics in an extensive simulation campaign.

% \begin{itemize}
% 	\item To the best of authors knowledge, this is the first work providing the definition of semantic transition model for a VSN, as well as the formulation of the semantic topology identification problem.
% 	\item The formalization of a suitable observation model and a network measurement generation protocol, according to the semantic transition model is presented.
% 	\item A methodology based on Deep Embedded Features to solve the proposed model identification problem is leveraged. More specifically, a Gaussian Mixture Model interpretation of the underlying semantic transition graph is proposed and an autoencoder is used to reduce the curse of dimensionality effects; then, semantic transition model is learnt by solving a Maximum A-Posteriori estimate on the encoded features.
% 	\item Specific and extensive numerical experiments are suggested to evaluate the proposed estimation algorithm.
% \end{itemize}

% PAPER STRUCTURE
\medskip
\noindent{\textbf{Paper structure - }}\label{subsec:paper_structure} Some mathematical preliminaries on the GMMs and the AEs are given in Section~\ref{sec:preliminaries}. Section~\ref{sec:problem} is devoted to the formal definition of the SM for a given VSN, and then to the statement the SM identification problem. Section~\ref{sec:methodology} focuses on the SM characterization, describing the stochastic models of the events and the network observations.
%describing the 
The outlined GMM-based SM identification solution is then illustrated in Section~\ref{sec:identification}.  Section~\ref{sec:numerical_results} reports the numerical results aiming at validating the designed procedure. The main conclusions are drawn in Section~\ref{sec:conclusion}.

% NOTATION
\medskip
\noindent{\textbf{Notation} -} Hereafter, lowercase italic characters denote
scalar values, lowercase bold characters denote 
(column) vectors, uppercase characters denote matrices, and uppercase calligraphic letters denote sets.
In addition, $\mathbf{1}_N \in \mathbb{R}^N$ and $\mathbf{0}_N \in \mathbb{R}^N$ respectively  identify the $N$-dimensional (column) vector of all ones and zeros, while $\mathbf{I}_{N\times N} \in \mathbb{R}^{N\times N}$ and $\mathbf{0}_{N\times N} \in \mathbb{R}^{N\times N}$ respectively 
identify the identity and null matrix of dimension $N$.\\
Furthermore, in this work, $\mathbb{SD}_{+}^N \subset \mathbb{R}^{N \times N}$ is the space of $N$-dimensional positive semi-definite matrices, $\mathbb{H}_u^N$ is the $N$-dimensional unitary hypercube,  $\mathbb{V}\mathbb{H}_u^{N}$ is the set of $2^N$ vertices of $\mathbb{H}_u^N$ and $\mathbb{V}\mathbb{H}_u^{N,0} = \mathbb{V}\mathbb{H}_u^{N}\setminus\{\mathbf{0}_N\}$ is the set of vertices of $\mathbb{H}_u^N$ excluding the origin (thus, its cardinality is equal to $2^N-1$).\\ 
Finally, the notation $\mathcal{N}(\mathbf{x}; \bm{\mu},\bm{\Sigma})$ is used to indicate the Gaussian distribution having parameters $\bm{\mu} \in \mathbb{R}^N$ and $\bm{\Sigma} \in \mathbb{R}^{N \times N}$ which best approximates the $N$-dimensional vector $\mathbf{x}$. Similarly, the notation $\mathcal{U}(y; a,b)$ is used in case of a uniform distribution with parameters $a,b \in \mathbb{R}$.

%%%%%%%%%%%%%%%%%%%%%%%%%%%%%%%%%%

% PRELIMINARIES
\section{{Methodological Background}}\label{sec:preliminaries}

% This section provides a mathematical background for the proposed SM identification procedure. Section~\ref{subsec:GMM_nearning} summarizes some principal foundations about Gaussian Mixture Models learning methods, while Section~\ref{subsec:AE} introduces the curse of dimensionality problem and how it can be eased.

This section provides a methodological background for the proposed SM identification procedure, revising the principal concepts related to the GMMs and the use of the auto-encoder to face the dimensionality problem.

% \subsection{Learning of Gaussian Mixture Models}\label{subsec:GMM_nearning}
\subsection{{GMMs Overview}}\label{subsec:GMM_learning}

%\LV{Read \cite{tzikas2008variational} to have an insight on EM-GMM and VB-GMM}\\
% Gaussian mixture models (GMM) are a valuable statistical tool
% for modeling densities, due to their flexiblity that allows to approximate any given density with high accuracy \cite{tzikas2008variational}. For this reason, GMMs have been widely used in a variety of applications ranging from speech understanding \cite{GMM_speech}, path planning \cite{GMM_search_static,GMM_search_RHC}, object tracking \cite{GMM_tracking}, image modeling \cite{GMM_image} and segmentation \cite{GMM_BS}. \\

Given the dataset $\mathcal{D} = \{\mathbf{x}^{d} \in \mathbb{R}^N \}_{d=1}^D$, the associated GMM is a parametric probability density function such that any data point $\mathbf{x}^{d} \in \mathcal{D}$ is statistically described by a weighted sum of %a suitable number $M \geq 1$ of 
Gaussian distributions.
Formally, introducing the weight set  $\{\pi_m  \in [0,1] \}_{m=1}^M$ having  cardinality $M \geq 1$ but such that $\sum_{m=1}^M \pi_m = 1$, it holds that
%let us introduce the latent or hidden \GMi{random} variable $z_m$ defined over the discrete alphabet $\mathcal{Z} = \{ z_m \in \mathbb{N} \}_{m=1}^{M}$ and such that
% \begin{equation}
% p(\vect{x}^d|z_m) = \mathcal{N} \left(\bm{\mu}_m, \bm{\Sigma}_m \right),
% \end{equation}
% where the vector $\bm{\mu}_m \in \mathbb{R}^N$ and the (positive semi-definite) matrix $\bm{\Sigma}_m \in \mathbb{R}^{N \times M}$ denote the unknown mean and unknown variance of the corresponding Gaussian distribution. From the total probability theorem, it follows that 
\begin{equation}
\label{eq:marginal_probability}
p(\vect{x}^d) = \sum_{m=1}^M \pi_m \; \mathcal{N} \left(\bm{\mu}_m, \bm{\Sigma}_m \right),
\end{equation}
where the vector $\bm{\mu}_m \in \mathbb{R}^N$ and the (positive semi-definite) matrix $\bm{\Sigma}_m \in \mathbb{R}^{N \times N}$ denote the unknown mean and unknown variance of the corresponding Gaussian distribution. 
%and any weight $\pi_m  \in [0,1]$ can be interpreted as the probability $\pi_m = p(z_m)$ of a certain (latent or hidden) \GMi{random} variable $z_m$ defined over the discrete alphabet $\mathcal{Z} = \{ z_m \in \mathbb{N} \}_{m=1}^{M}$, with unknown cardinality $M$.
% \begin{equation}
%     \pi_m = p(z_m) \in [0,1], \quad \sum_{m=1}^M \pi_m = 1.
% \end{equation}
%

The $N$-dimensional GMM having $M$ components and corresponding to the probability density function~\eqref{eq:marginal_probability} is thus parametrized by the unknown set $\bm{\phi} = \{ \bm{\phi}_m \}_{m=1}^M$ where
\begin{equation}
\bm{\phi}_m = \left\lbrace \pi_m, \bm{\mu}_m, \bm{\Sigma}_m \right\rbrace \in \Phi \subseteq [0,1] \times \mathbb{R}^N \times \mathbb{SD}_{+}^N.
\end{equation}
%In the light of this fact, hereafter, the notation $p_{\bm{\phi}}(\cdot) =p(\cdot)$ is used.
%, namely
% \begin{equation}
% \mathbb{S}_{+}^N = \left\lbrace \vect{A} \in \mathbb{R}^{N \times N} : \vect{A}=\vect{A}^\top, \; \vect{A} \curlyeqsucc \vect{0}  \right\rbrace.
% \end{equation}
{In the literature, it exists a collection of techniques aiming at identifying the optimal parameter set $\bm{\phi}^\star$ whose corresponding  probability density function better fits with the given dataset $\mathcal{D}$. These are generally referred to as \textit{GMM learning methods} and distinguished into two classes according to the adopted Fisherian or Bayesian approach. In the former case, the learning strategy rests on the solution of a maximization problem, i.e., the Maximum Likelihood Estimation (MLE) problem, by exploiting an iterative procedure, known as Expectation-Maximization (EM) algorithm, which outputs an estimate of the optimal parameter set $\bm{\phi}^\star$. On the other side, when resting on \textit{Bayesian approach}, the parameters estimation is replaced by the inference of their a-posteriori probability distributions. In particular, the most common \textit{Variational Bayesian (VB)} techniques exploit some auxiliary hyperparameters to characterize the a-posteriori distribution of the optimal parameters set $\bm{\phi}^\star$~\cite{tzikas2008variational,li2018clustering}. %distributedEM,VariationalGMM,
}
 {In addition, when resting on a Fisherian approach, the well-known Cross-Validation methods can be employed to determine also the optimal number $M^\star$ of components of the considered GMM. Contrarily, the Variational inference method is such that some mixture weights are selected close to zero, thus suggesting the number of effective components of the GMM~\cite{li2018clustering}.}

\subsection{AE-based dimensionality reduction}\label{subsec:AE}

Any GMM learning method can be interpreted as a \textit{soft-clustering strategy}~\cite{DEC}, having the purpose of assigning any data point $\mathbf{x}^d \in \mathcal{D}$ to the $m^\star$-th cluster with 
\begin{equation}\label{eq:GMM_clustering}
m^\star = \argmax_{m \in [1,M]}{ \frac{\pi_m \; \mathcal{N}(\bm{\mu}_m,\bm{\Sigma}_m)}{\sum_{k=1}^M \pi_k \; \mathcal{N}(\bm{\mu}_k,\bm{\Sigma}_k) }}.
\end{equation}
In the light of this fact, one can observe that, like any other clustering algorithm that aggregates data with similar properties, criticalities may emerge when accounting for GMMs in high-dimensional spaces~\cite{steinbach2004,dimensionalityReduction_tutorial, dimensionalityReduction_review}.
%,bellman1966dynamic

\begin{figure}[t!]
\centering
\includegraphics[width=0.85\columnwidth]{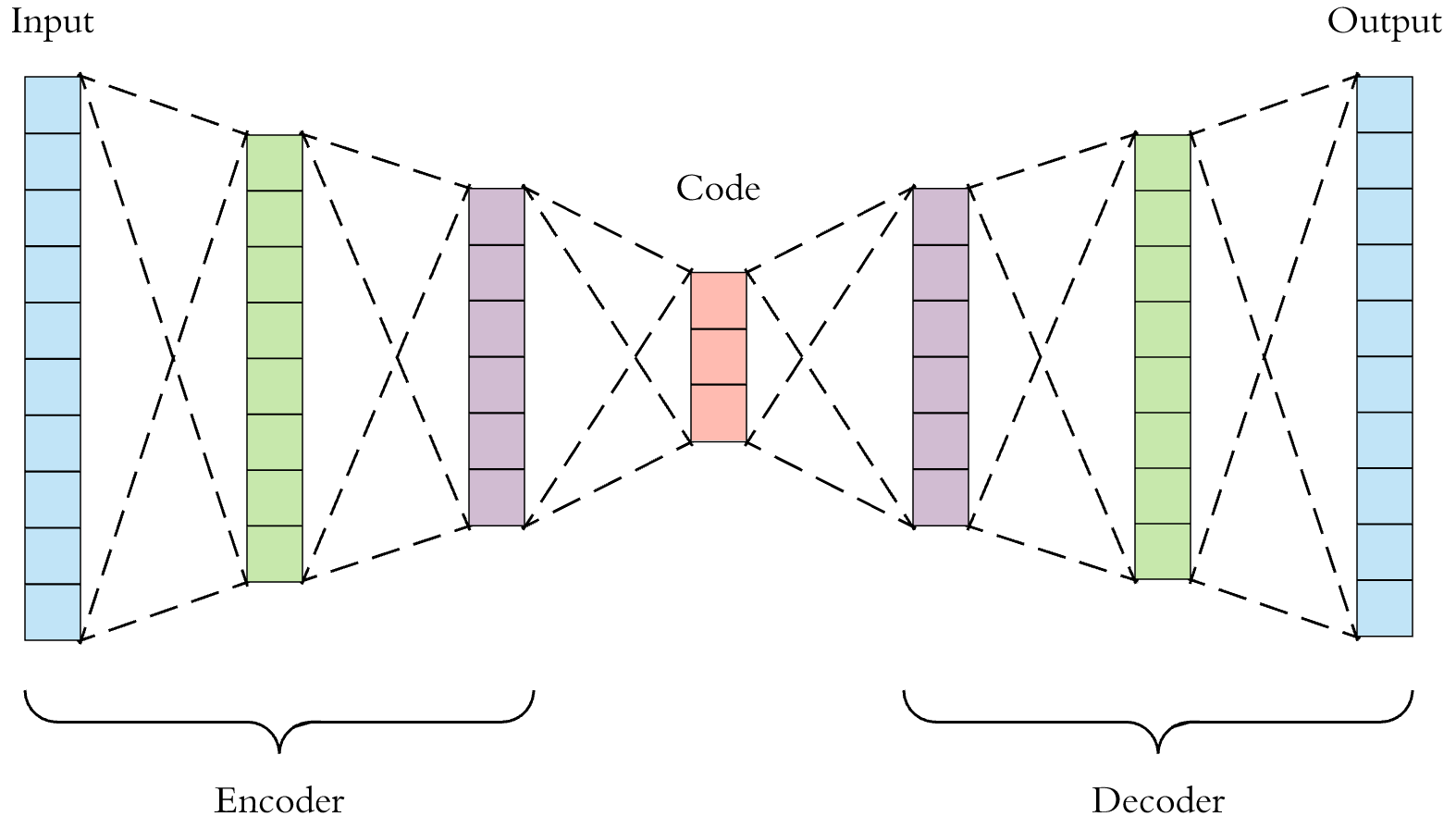}
\caption{\footnotesize
Typical AE structure.}
\label{fig:AE}
\end{figure}

To face this drawback, an AE can be exploit to carry out a dimensionality reduction, thus entailing the overfitting abatement, the data interpretability increase and the required storage space reduction.  
% In the literature, several dimensionality reduction techniques are known~\cite{dimensionalityReduction_review}. %dimensionalityReduction_tutorial,
% {Among them, the solutions based on the AEs (see, e.g.~\cite{charte2018practical}) %wang2016auto,DL_tutorial, 
% are characterized by well-stated theoretical function approximation properties and demonstrated learning capabilities since these are based the deep neural networks~\cite{bengio2013representation}.} %hornik1991approximation, %Furthermore, AEs are proven to learn data projections that are more interesting than other basic techniques \cite{wang2016auto}.\\
An AE is an artificial neural network designed to learn a dataset representation in an unsupervised manner~\cite{wang2016auto,DL_tutorial, charte2018practical}. As reported in Figure~\ref{fig:AE}, this consists of two parts, the \textit{encoder} and the \textit{decoder}. Introducing the \textit{input set} $\mathcal{X} \subseteq \mathbb{R}^d$ so that $\mathcal{D} \subseteq \mathcal{X}$ and the \textit{feature set}  $\mathcal{F} \subseteq \mathbb{R}^p$, with $p < d$, the encoder and the decoder  can be defined via the  functions $f^\star: \mathcal{X} \rightarrow \mathcal{F}$ and $g^\star: \mathcal{F} \rightarrow \mathcal{X}$ such that
\begin{align}
(f^\star,g^\star) = \argmin{\lVert \mathcal{D} - g( f(\mathcal{D}) ) \rVert^2}.
\end{align}
%Note that %, since the feature space $\mathcal{F}$ have lower dimensionality than the input space $\mathcal{X}$,
Any vector $f (\vect{x}) \in \mathbb{R}^p$ can be interpreted as a compressed representation of the vector $\vect{x} \in \mathbb{R}^d$, however a key property of the AEs is that the non-linear transformation $f(\cdot)$ preserves local data structure% the structure of the input data in terms of probability distribution
~\cite{goodfellow2016deep}. %CAE,peng2016deep, %:  this is of paramount importance in clustering tasks, where similarity properties between features are used for data aggregation.

%%%%%%%%%%%%%%%%%%%%%%%%%%%%%%%%%%

% PROBLEM STATEMENT
\section{VSN Stimulation Model}\label{sec:problem}

\begin{figure*}[t!]
\centering
\subfloat[\label{fig:multigraph_DEI}]{
\includegraphics[scale=0.3]{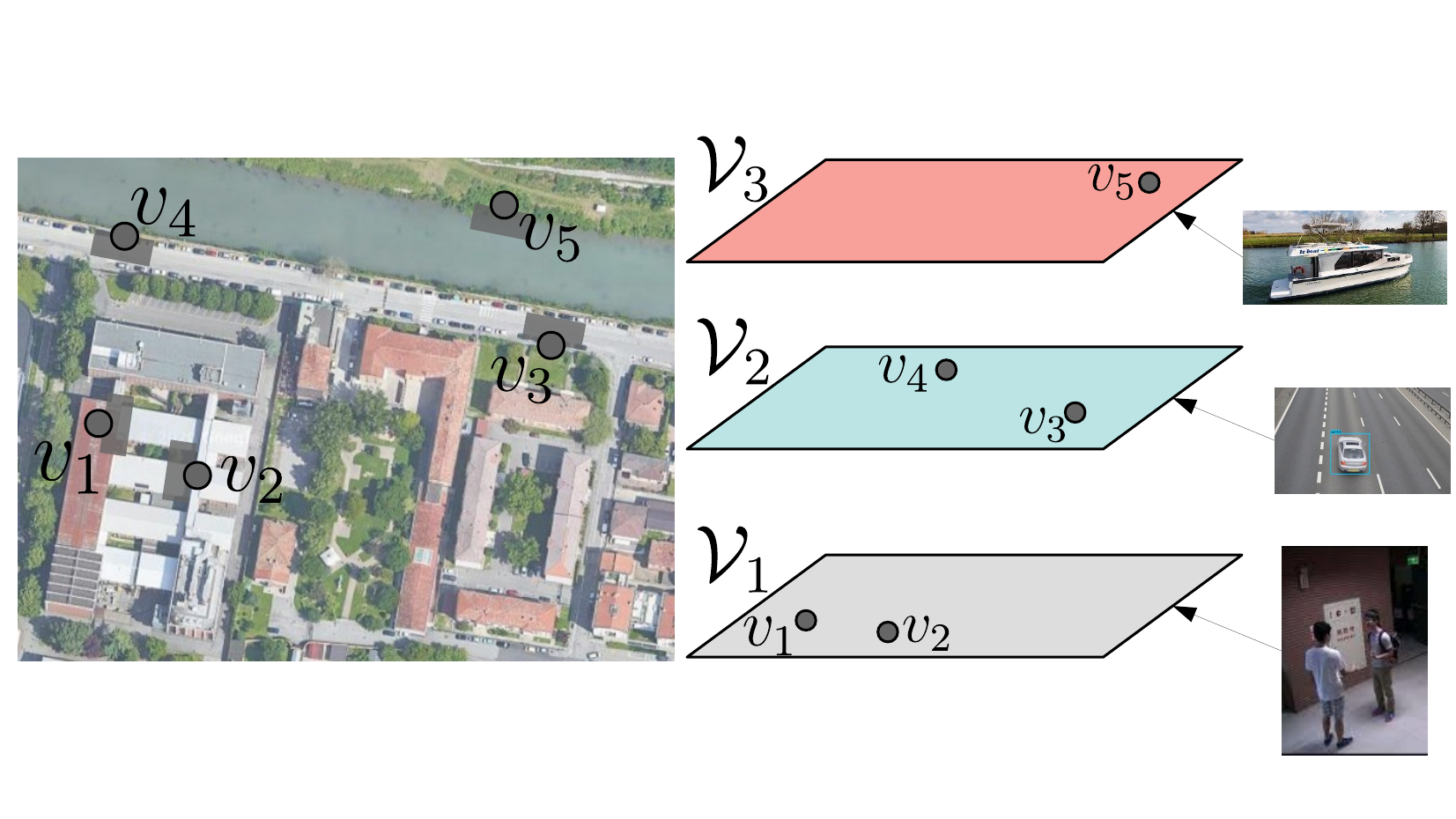}
}%
~
\subfloat[\label{fig:multigraph_warehouse}]{
\includegraphics[scale=0.3]{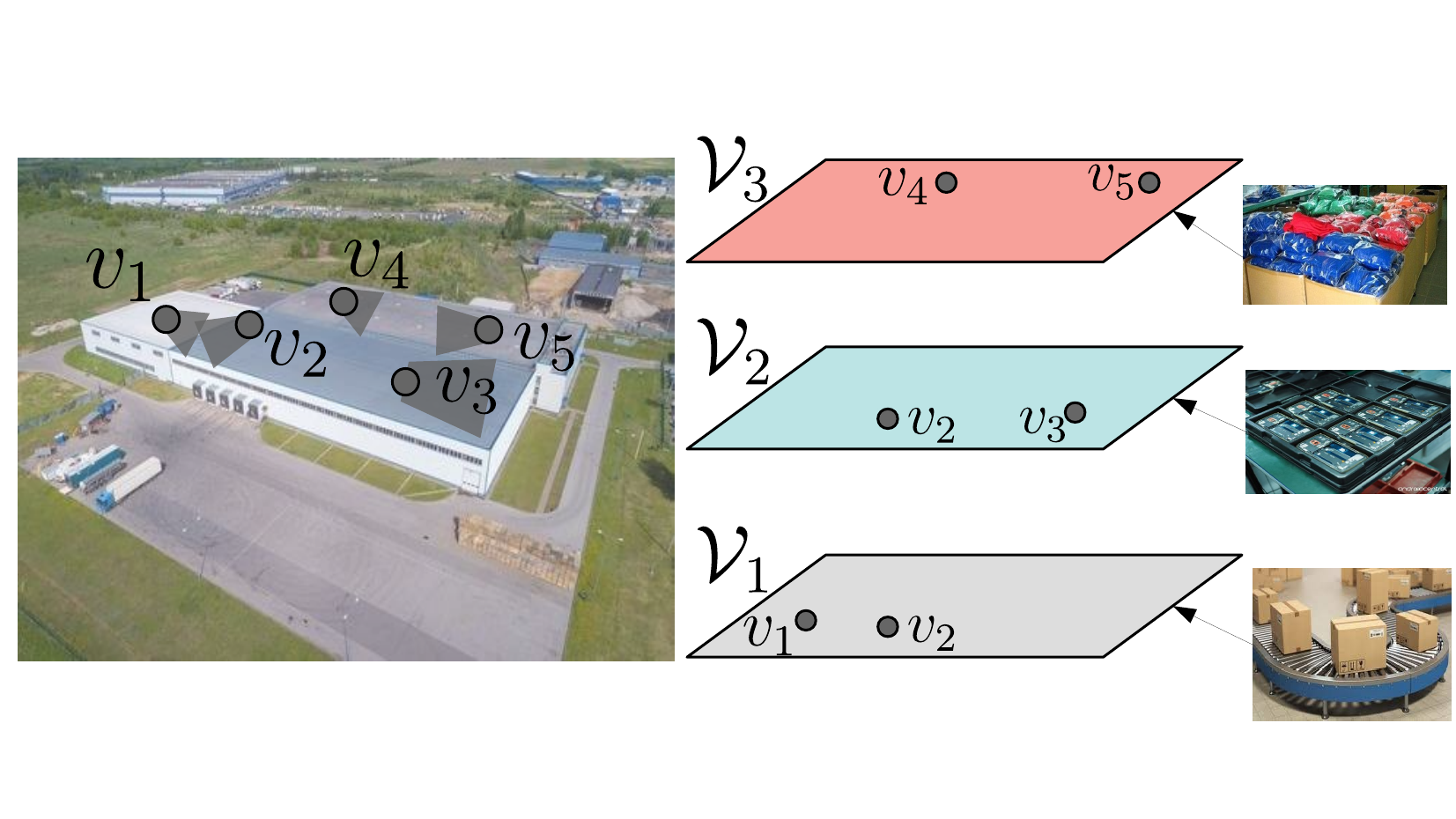}}
\caption{Examples of SMs in real-world scenarios wherein $M=3$ events stimulate a VSN composed of $N=5$ cameras placed in urban \textbf{(a)} and industrial context \textbf{(b)} - gray markers represent the  cameras, while the gray-shaded areas denote their FoV. }
\label{fig:multigraph}
\end{figure*}%

This section is devoted to the statement of the \textit{SM identification problem} for a given VSN. The attention is focused on the definition of the SM associated to the multi-camera system, and then on the formalization of the identification problem in terms of estimation of a suitable Boolean matrix.

% Section~\ref{subsec:camera_network} defines the SM and clarifies the relation with the VSN semantic sensing topology. In Section~\ref{subsec:stimulation} the definition is reformulated so that the identification problem is cast into a Boolean matrix estimation task.

\subsection{SM Definition}\label{subsec:camera_network}

Let us consider a VSN made up of $N \geq 2$ cameras, and 
a set of possible \textit{events} \mbox{$\mathcal{E}\!=\!\{ e_m \}_{m=1}^{\overline{M}},\; \overline{M} \!\in \![1,2^N-1]$}.  In this work, the term \textit{event} is used to indicate a generic occurrence that a camera can detect through the state-of-the-art computer vision techniques, as for instance, the presence of a specific object or person, but also a pre-defined action, gesture or behavior performed by one or more agents.  
Although any camera in the given network can be programmed to detect all the events in $\mathcal{E}$~\cite{ahmad2019deep}, % objectDetection_survey,actionRecognition_survey,
only a subset $\mathcal{E}_a = \{e_m\}_{m=1}^M \subseteq \mathcal{E}$, $M \leq \overline{M}$, of events are practically perceivable by the VSN devices. Hereafter, these are referred to as \emph{active events}, and for sake of completeness, the events belonging to the set  $\mathcal{E}_s = \mathcal{E} \setminus \mathcal{E}_a$ are indicated as \textit{silent events}. 

\begin{example}
To clarify this fact, let us consider a VSN whose cameras are programmed to detect and classifying moving agents, including both cars and people. Whether the network is employed in indoor environment, any vehicle occurrence constitutes a silent event, whereas walking children represent active events.
\end{example}

Accounting for the graph-based multi-element systems representation, 
let us introduce the node set $\mathcal{V} = \{v_n\}_{n=1}^N$  such that the $n$-th node $v_n$ represents the $n$-th device in the network. The occurrence of any active event $e_m \in \mathcal{E}_a$ inside the FoV of the $n$-th camera is, thus, referred to as the \emph{stimulation} of $v_n$ by $e_m$. Moreover, it is possible to define the VSN  \textit{SM} as follows. 

\begin{definition}\label{def:stimulation_model}
For a $N$-devices VSN associated to the node set  $\mathcal{V} = \{v_n\}_{n=1}^N$, and the corresponding set of active events $\mathcal{E}_a = \{e_m\}_{m=1}^M$, the \textit{SM} consists in the function
\begin{equation}\label{eq:selective_function}
\tau : \;
\mathcal{E}_a \rightarrow \mathbb{P} (\mathcal{V}), \;\; e_m \mapsto \mathcal{V}_m,
\end{equation}
where $\mathbb{P} (\mathcal{V})$ denotes the power set of $\mathcal{V}$ and $\mathcal{V}_m$ identifies the set of cameras stimulated by the event $e_m$, namely the so-called $m$-th stimulation set.
\end{definition}

According to Definition~\ref{def:stimulation_model}, the VSN SM can be mathematically represented by the multi-graph $\mathcal{G}^\tau= \{ \mathcal{G}_{m} = (\mathcal{V}_m, \mathcal{V}_m \times \mathcal{V}_m) \}_{m=1}^M$, resulting in a collection of undirected and fully connected graphs, hereafters referred to as \textit{communities}~\cite{jin2020identification,wang2019temporal}. %,huang2019community
On the other hand, as concerns function~\eqref{eq:selective_function}, the following remarks are mandatory.

\begin{rem}\label{R1}
Since two distinct events can stimulate the same set of cameras,
the function $\tau(\cdot)$ is not guaranteed to be injective, namely $e_m,e_{m'} \in \mathcal{E}_a$ with $e_m \neq e_{m'}$ generally does not imply $\mathcal{V}_m \neq \mathcal{V}_{m'}$. Thus, it follows that identical communities can be present in $\mathcal{G}^\tau$.  
\end{rem}

% From Remark~\ref{R1}, it follows that identical communities can be present in $\mathcal{G}^\tau$.

\begin{rem}\label{R2}
By definition of active events, any stimulation set $\mathcal{V}_m$ cannot be an empty set, and, therefore, it holds that $\emptyset \notin \Ima \tau (\mathcal{E}_a)$. This implies that the function $\tau(\cdot)$ is neither surjective. In particular, in general it also holds that $\vert \Ima \tau (\mathcal{E}_a) \vert = M < 2^N = \vert \mathbb{P}(\mathcal{V}) \vert$. 
\end{rem}

\begin{rem}\label{R3}
Despite $\bigcup_{m=1}^M \mathcal{V}_m = \mathcal{V}$, the stimulation sets are not guaranteed to be pairwise disjoint, since the same camera can be stimulated by different events. Thus, the communities of $\mathcal{G}^\tau$ do not represent clusters of the VSN~\cite{LuvisottoClustering}. %,schaeffer2007graph
\end{rem}

The following example is provided to clarify the aforementioned problem setup.

\begin{example}
Figure~\ref{fig:multigraph} provides two examples of SMs accounting for real-world scenarios, both characterized by $M=3$ active events and a VSN composed of $N=5$ cameras. 
In Figure~\ref{fig:multigraph_DEI} a urban scenario is considered. The camera-nodes $v_1$ and $v_2$ are placed inside a building, the camera-nodes $v_3$ and $v_4$ monitor a road, the camera-node $v_5$ is positioned along a river. In addition, the event $e_1$ corresponds to the detection of a person, the event $e_2$ to the detection of a vehicle and the event $e_3$ to the detection of a boat. According to the collected data, it is possible to distinguish the following stimulation sets: $\mathcal{V}_{1} = \{v_1,v_2\}$, $\mathcal{V}_{2} = \{v_3,v_4\}$ and $\mathcal{V}_{3} = \{v_5\}$. 
In Figure~\ref{fig:multigraph_warehouse} an industrial warehouse is depicted. This is composed of three structured rooms: one for generic packages storage (corresponding to event $e_1$) monitored by the camera-nodes $v_1$ and $v_2$, one for mobile phones storage (corresponding to event $e_2$) surveilled by camera-nodes $v_2$ and $v_3$, and one for clothes storage (corresponding to event $e_3$) wherein the camera-nodes $v_4$ and $v_5$ are located. In this case, the stimulation sets result to be $\mathcal{V}_{1} = \{v_1,v_2\}$, $\mathcal{V}_{2} = \{v_2,v_3\}$ and $\mathcal{V}_{3} = \{v_4,v_5\}$.
\end{example}

\subsection{SM Identification Problem}\label{subsec:stimulation}

According to Definition~\ref{def:stimulation_model}, the SM for a $N$-devices VSN consists in the function~\eqref{eq:selective_function}, which maps any active event $e_m \in \mathcal{E}_a$ in the corresponding stimulation set $\mathcal{V}_m \in \mathbb{P}(\mathcal{V})$, $m \in [1,M]$.
Nonetheless, it is to be noted that the aforementioned function $\tau(\cdot)$ is equivalent to the following function
\begin{equation}\label{eq:stimulation_model}
\tau^\prime: \; [1,M] \rightarrow \{0,1\}^N, \quad m \mapsto \vect{t}_m,
\end{equation}
where the index $m \in [1, M]$ uniquely identifies the $m$-th event in $\mathcal{E}_a$ and the vector $\vect{t}_m \in \{0,1\}^N$ is a $N$-dimensional binary representation of the stimulation set $\mathcal{V}_m$, hereafter referred to as \textit{stimulation vector}. In detail, each $n$-th entry of $\vect{t}_m$, $n \in [1, N]$, is defined as follows 
\begin{equation}\label{eq:T_m}
t_{m,n}=
\begin{cases}
1, \quad v_n \in \mathcal{V}_m \\
0, \quad \text{otherwise}
\end{cases}.
\end{equation}

Through the introduction of function~\eqref{eq:stimulation_model}, the SM is completely characterized by the so-called \textit{stimulation matrix}, namely the Boolean matrix obtained by stacking all the stimulation vectors. Formally, this is given by \begin{equation}\label{eq:stimulation_matrix}
\vect{T} = \begin{bmatrix}
\vect{t}_1 & \dots & \vect{t}_M
\end{bmatrix}^\top \in \{0,1\}^{M \times N}.
\end{equation}
Note that each row of the matrix $\vect{T}$ corresponds to an event in $\mathcal{E}_a$. The following problem can be thus stated.

\begin{problem}
\label{prob:stimulation_model_identification}
For a $N$-devices VSN associated to the node set  $\mathcal{V} = \{v_n\}_{n=1}^N$ and the corresponding set of active events $\mathcal{E}_a = \{e_m\}_{m=1}^M$, the \textit{SM identification problem} consists in the estimation of the stimulation matrix $\vect{T} \in \{0,1\}^{M \times N}$, defined according  to~\eqref{eq:stimulation_model}-\eqref{eq:stimulation_matrix}, up to row permutations.
\end{problem}

Note that, for a $N$-devices VSN, 
% the maximum number of different and detectable active events is equal to $2^{N}$, whereas 
Problem~\ref{prob:stimulation_model_identification} results to be well defined when $M \leq 2^{N}$. In fact, the condition $M > 2^{N}$ implies the existence of at least two different indistinguishable events which are associated to the same stimulation vector. On the other hand, from Remark~\ref{R2} it follows that $\vect{0} \notin \Ima \tau^\prime([1,M])$, and hence it is not possible to have $M = 2^{N}$.   
\section{SM Characterization}
\label{sec:methodology}

To subsequently cope with the SM identification problem in a formal manner, this section focuses on the adopted stochastic description of both the active events and the network observations understood as the stimulation vectors generated from the cameras collected data.

\subsection{Active Events}\label{subsec:prior}

In general, the occurrence of an event in a certain monitored environment
can be described by the realization of a random variable.  
Accounting for the set $\mathcal{E}_a$, any active event $e_m$, $m \in [1, M]$ can be modeled as a discrete random variable, whose corresponding probability distribution $p(e_m) \in [0,1]$ represents the (unknwon) \mbox{a-priori} probability of the event to occur. In particular, without loss of generality, it can be assumed that $ \sum_{m=1}^M p(e_m) = 1.$ %\begin{equation}\label{eq:a_priori}
% \sum_{m=1}^M p(e_m) = 1.
% \end{equation}

In addition, the following further assumption is made as concern the active events. This will turn out to be relevant within the observations generation (Section~\ref{subsec:observation}).

\begin{assumption}\label{assumption:trajectory}
	When any event $e_m$, $m \in [1, M]$ occurs in the monitored environment according to the given stochastic law, it appears at least once in the FoV of all and only the cameras within the corresponding stimulation set $\mathcal{V}_m$, defined according to~\eqref{eq:selective_function}.
\end{assumption}

\subsection{Network Observations}\label{subsec:observation}

To face Problem~\ref{prob:stimulation_model_identification}, it is necessary to adopt a model for the network observations able to describe the stimulation vectors generated by the collected camera data, by also taking into account misdetections and misclassifications. 

In doing this, let us introduce the observation dataset  \begin{equation}\label{eq:dataset_network}
	\mathcal{D} = \left\lbrace \widetilde{\vect{t}}^{d} \in \mathbb{H}_u^N \right\rbrace_{d=1}^D.
\end{equation}
In~\eqref{eq:dataset_network} any $d$-th vector $\widetilde{\vect{t}}^{d} \in \mathbb{H}_u^N$ denotes the stimulation vector, %corresponding to the $m$-th event, $m \in [1,M]$, 
obtained through the aggregation of the data collected by the single cameras as reported in Algorithm~\ref{algo:measurement_protocol}. 

According to Algorithm~\ref{algo:measurement_protocol}, when a $n^\prime$-th camera, $n^\prime \in [1,N]$ identifies the event $e_m$, $m \in [1,M]$, it generates the corresponding observation $\widetilde{t}_{n^\prime} = c_{n^\prime}$,  where $c_{n^\prime} \in [0,1]$ represents the confidence index associated to the event recognition. According to the literature~\cite{yolo}, %,hung2020faster
such an index  is supposed to be extracted from a uniform distribution $\mathcal{U}(c;\underline{c},1)$ where $\underline{c}$ is typically chosen close to $1$.
Then, under the assumption of perfect network communication, the camera-node $v_{n^\prime}$ notifies to all the other network components about the occurrence of the $m$-th event in the monitored environment\footnote{For a camera, the \textit{event identification} (recognition) consists in the event detection and classification, while the \textit{stimulation} coincides with the occurrence of an event inside its FoV. The former process can be subject to misclassifications, the latter to both misdetections or misclassifications.}. According to Assumption~\ref{assumption:trajectory}, the event $e_m$ appears (at least once) inside the FoV of all and only the cameras in {$\mathcal{V}_m \setminus \{v_{n^\prime}\}$}. However, all the VSN devices wait for the event for $K\geq 1$ iterations, where the \textit{patience parameter} $K$ constitutes a tunable variable. At each $k$-th iteration, $k \in [1,K]$, any $n$-th camera in {$\mathcal{V} \setminus \{v_{n^\prime}\}$} updates the value of its observation $\widetilde{t}_{n} \in [0,1]$ as follows
\begin{equation}\label{eq:camera_measurement}
\widetilde{t}_{n} = h(m,n)c_n,  \;
	h(m,n) = 
	\begin{cases}
		1, \text{ if } v_n \text{ identifies } e_m\\
		0, \text{ otherwise}.
	\end{cases}
\end{equation}
Finally, the observation vector $\widetilde{\vect{t}}$ is obtained by stacking the outputs of all the VSN components; hence, it is $\widetilde{\vect{t}} \in \mathbb{H}_u^N$.  

\begin{algorithm}[t!]
	\small
	
	\SetKwInOut{Input}{input}
	\textbf{initialization:} 
	\begin{itemize}[noitemsep]
		\item the camera-node $v_{n^\prime}$ identifies the event $e_m$
		\item the camera-node $v_{n^\prime}$ computes  $\widetilde{t}_{n^\prime} = c_{n^\prime}$
		\item the camera-node $v_{n^\prime}$ communicates to all other devices
	\end{itemize}
	\For{$v_n \in \mathcal{V} \setminus v_{n^\prime}$}{
		\vspace{0.2cm}
		$\widetilde{t}_{n} = 0$\\
		$k = 1$ \\
		\While{$k \leq K$}{
			\uIf{$h_{m,n} =1$ }{
			\vspace{0.2cm}
				$\widetilde{t}_{n}= c_n$ \\
				break
			}
			\Else{
			    $k \leftarrow k+1$
			}
		}
	}
	$\widetilde{\vect{t}} = \begin{bmatrix} \widetilde{t}_{1} & \dots & \widetilde{t}_{N} \end{bmatrix}^\top  \in \mathbb{H}_u^N$
	\caption{\footnotesize Network Observations Generation}
	\label{algo:measurement_protocol}
\end{algorithm} 

In~\eqref{eq:camera_measurement}, $h(m,n)$ plays the role of an indicator function that is active (i.e., $h(m,n)=1$) only when the $n$-th camera \textit{identifies} the event $e_m$, namely in correspondence to one of the following cases: $(a)$ the event $e_m$ stimulates the camera-node $v_n$, which correctly classifies the event $e_m$; $(b)$ the event $e_{m^\prime}, \; m^\prime \neq m$, stimulates the camera-node $v_n$, which incorrectly classifies the event $e_{m^\prime}$ as $e_{m}$. 
The latter case is common when event recognition algorithms are fooled by objects or actions described by similar features~\cite{yoon2019structural}. Conversely, it holds that $h(m,n)=0$ when either the event $e_m$ does not stimulates the camera-node $v_n$, or when the camera-node $v_n$ misdetects the event $e_m$, e.g., due to occlusions and/or low resolution issues~\cite{sanmiguel2017efficient}.

To capture the complexity and the probabilistic nature of the described  scenario, it turns out to be advantageous to introduce a stochastic characterization of the indicator function $h(\cdot,\cdot)$. In doing this, the conditional probability of $h(m,n) =1$ given that the event $e_m$ occurs in the monitored environment can be computed as follows
{\setlength{\mathindent}{0.2cm}
\begin{align}
\label{eq:h_i_1}
 p_{m,n}^{h1|m} 
&=p\left(h(m,n) =1 | e_m \right) = 
t_{m,n} 
p(D_{m,n}) p(C_{m \rightarrow m,n}) + \nonumber \\
& \qquad	 + \sum_{\substack{m^\prime \neq m \; : \\ \; v_n \in \mathcal{V}_{m^\prime}}}p(S_{m^\prime,n}) p(D_{m^\prime,n}) p(C_{m \rightarrow m^\prime,n})  \\
	p_{m,n}^{h0|m} &=	p\left(h(m,n) =0 \vert e_m \right) = 1 - p_{m,n}^{h1|m}
\end{align}
}
In~\eqref{eq:h_i_1} $p(S_{m,j}) \in [0,1]$ denotes the probability that event $e_m$ stimulates camera $v_n$. Thus, it is  
\begin{equation}\label{eq:p_stimulation}
\begin{split}
    p(S_{m,n}) 
	& = \begin{cases}
		p(e_m), & \text{ if } v_n \in \mathcal{V}_m \\
		0, & \text{ otherwise}.
	\end{cases}
	\end{split}
\end{equation}
The probability that the $n$-th camera detects the $m$-th event is instead represented by $p(D_{m,n}) \in [0,1]$, while $p(C_{m \rightarrow m,n}) \in [0,1]$ ($p(C_{m \rightarrow m^\prime,n}) \in [0,1]$) is the probability that the $n$-th camera classifies the detected event as $e_m$ (as $e_{m^\prime}$). 

For sake of simplicity,  
hereafter, it is assumed that
	\begin{align}
		& p(D_{m,n}) = p_D \label{eq:p_D_constant} \\
		& p(C_{m \rightarrow m,n}) = p_C \label{eq:p_C_constant} \\
		& p(C_{m \rightarrow m^\prime,n}) = \frac{1 - p_C}{\overline{M}-1},  \label{eq:p_misclassification}
	\end{align}
for any $m \in [1,M]$, $m^\prime \in [1,\overline{M}]$ and $n \in [1,N]$.
In other words, the detection probability~\eqref{eq:p_D_constant} and correct classification probability~\eqref{eq:p_C_constant} are supposed to be independent on both the considered event and  camera (i.e., there is no dependency on the indices $m$ and $n$). In particular, $D_{m,n}$ is a Bernoulli random variable with success probability $p_D$; whereas the misclassification probability~\eqref{eq:p_misclassification} of any event is equally distributed among all the  $\overline{M}-1$ events (excluding $e_m$) 
%\LV{(i.e., $C_{m \rightarrow m^\prime,n}$ is a uniform random variable over $\overline{M}-1$ elements)}.  
Under these assumptions, it holds that
\begin{equation}\label{eq:h_i_1_reformulate}
		p_{m,n}^{h1|m}
		\!=\! p_D \left( t_{m,n}p_C + \frac{1 \!-\! p_C}{\overline{M}\!-\!1} \sum_{m^\prime \neq m} \! p(e_{m^\prime})t_{m^\prime,n}  \right) \end{equation}
According to~\eqref{eq:h_i_1_reformulate}, %for any $m \in [1,M]$ and {$n \in [1,N]$}, 
the probability of the $m$-th event identification by the $n$-th camera results from three contributions: the event detection probability $p_D$ which works as necessary condition for $v_n$ to identify $e_m$, the probability $p_C$ of correct classification of $e_m$ by the $n$-th camera, and a combination of the cumulative a-priori probability of any other event (sum over the events different from $e_m$ and able to stimulate $v_n$) and the misclassification probability~\eqref{eq:p_misclassification}. 

Given these premises, it is possible to define the 
probability distribution of any camera observation over the $K$ iterations, given the occurrence of the $m$-th event in the environment.
% probability distribution of any single camera to be stimulated by the $m$-th event in the next $K$ iterations. 
This, called 
\emph{conditional single camera stimulation likelihood}, 
turns out to be
{\setlength{\mathindent}{0.2cm}
\begin{align}
		 p(\widetilde{t}_{n}\vert e_m)
		& = \left( 1-(p_{m,n}^{h0|m})^K \right){\mathcal{U}(\widetilde{t}_{n};\underline{c},1)} + (p_{m,n}^{h0|m})^K\delta(\widetilde{t}_{n}) \nonumber \\
		& = (1-\alpha_{m,n}){\mathcal{U}(\widetilde{t}_{n};\underline{c},1)} + \alpha_{m,n}\delta(\widetilde{t}_{n}) \label{eq:single_camera_likelihood}
\end{align}}
where $\delta(\cdot)$ is the Kronecker delta and  $\alpha_{m,n} = (p_{m,n}^{h0|m})^K$ is introduced to ease the notation. Notice that the \textit{incorrect observation probability} corresponds to $\alpha_{m,n}$ if $t_{m,n}=1$ and to  $1-\alpha_{m,n}$ if $t_{m,n}=0$. In detail, in the former case, the systematic error affecting the correct observations is equal to $(\underline{c}+1)/2$, which is thus dependent from the (minimum) confidence of the event recognition algorithm.

\begin{rem}
The probability of detecting the $m$-th event by the $n$-th camera is proportional to the parameter $K$. Thus, when $v_n \in \mathcal{V}_m$, a high value of $K$ results to be beneficial. However, this choice increase the possibility of artifacts emergence in the observations. As a consequence, it is suitable to select the patience parameter by accounting for the specific application, the cameras computational capabilities and placement and the structure of the monitored environment.
\end{rem}

Given~\eqref{eq:single_camera_likelihood}, it is possible to define also the \emph{conditional VSN stimulation likelihood} as the multivariate $N$-dimensional distribution $p(\widetilde{\vect{t}}\vert e_m) = p(\widetilde{t}_{1},\dots,\widetilde{t}_{N}\vert e_m)$.
Assuming i.i.d. cameras observations, it holds that
{\setlength{\mathindent}{0.2cm} \small
\begin{align}\label{eq:network_likelihood_iid}
p(\widetilde{\vect{t}}\vert e_m)
    	& = \prod_{n=1}^N p(\widetilde{t}_{n}\vert e_m)  \nonumber \\
    	& = \prod_{n=1}^N \left( (1-\alpha_{m,n})\mathcal{U}(\widetilde{t}_{n} ; \underline{c},1) + \alpha_{m,n}\delta(\widetilde{t}_{n}) \right) \nonumber\\
    	& \!=\! {\small \sum_{k_1=0}^1 \!\! \dots \!\! \sum_{k_N=0}^1 \left((1-\alpha_{m,1})\mathcal{U}(\widetilde{t}_{1} ; \underline{c},1)\right)^{k_1} \! \left( \alpha_{m,1}\delta(\widetilde{t}_{1})\right)^{1-k_1} \cdots} \nonumber \\
    	&\quad  \! {\small \cdots  \left((1-\alpha_{m,N})\mathcal{U}(\widetilde{t}_{N} ; \underline{c},1)\right)^{k_N}\left( \alpha_{m,N}\delta(\widetilde{t}_{N})\right)^{1-k_N}}. 
\end{align}
}

% \begin{equation}
% \begin{split}
%     	p(\widetilde{\vect{t}}_m) 
%     	& = \prod_{i=1}^N p(\widetilde{t}_{m,i}) \\
%     	& = \prod_{i=1}^N \left[ (1-\alpha_{m,i})\mathcal{U}(\widetilde{t}_{m,i}|c,1) + \alpha_{m,i}\delta(\widetilde{t}_{m,i}) \right] \\
%     	& = \sum_{k_1=0}^1 \!\! \dots \!\! \sum_{k_N=0}^1 \left[(1-\alpha_{m,1})\mathcal{U}(\widetilde{t}_{m,1}|c,1)\right]^{k_1} \! \left[ \alpha_{m,1}\delta(\widetilde{t}_1)\right]^{1-k_1} \times \\
%     	& \dots \times \left[(1-\alpha_{m,N})\mathcal{U}(\widetilde{t}_{m,N}|c,1)\right]^{k_N}\left[ \alpha_{m,N}\delta(\widetilde{t}_{m,N})\right]^{1-k_N}.
% \end{split}
% \end{equation}

\begin{figure}[t!]
	\centering
	\includegraphics[scale=0.32]{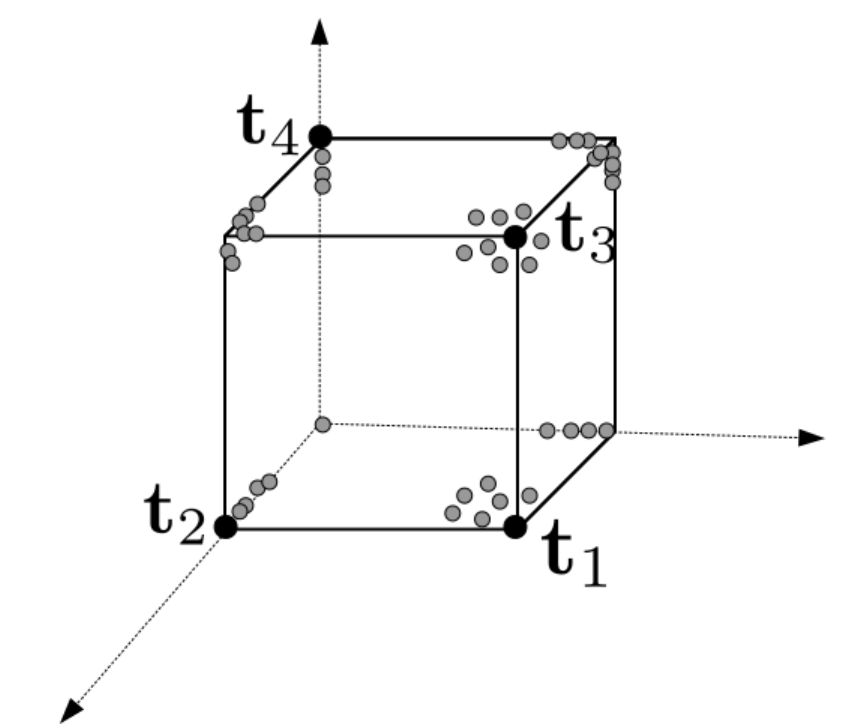}
	\caption{\footnotesize
		Visual representation of the geometrical interpretation of the stimulation process, with $N=3$ and $M=4$. Stimulation vectors $\vect{t}_m$ (larger black markers) are on the vertices of the unit hypercube of dimension $3$ ($\mathbb{H}_u^3$); the network observations $\widetilde{\vect{t}}_m$ (smaller gray markers) live inside $\mathbb{H}_u^3$, according to~\eqref{eq:single_camera_likelihood}.}
	\label{fig:cube}
\end{figure}

\begin{rem}
\label{rem:geometric_interpretation}
\textbf{\emph{(Geometric interpretation)}}
According to~\eqref{eq:selective_function}, each event $e_m$ is associated to a stimulation set $\mathcal{V}_m$, and, based on~\eqref{eq:stimulation_model}, such a set can be described by the $N$-dimensional binary vector $\vect{t}_m$. Thus, 
any vector $\vect{t}_m$, can be interpreted as an element of $\mathbb{V}\mathbb{H}_u^{N,0}$. In particular, the set $\lbrace \vect{t}_m \rbrace_{m=1}^M$ identifies the \emph{active vertices}. Then, in accordance with~\eqref{eq:single_camera_likelihood}, each component $\widetilde{t}_{n}$ of the observation vector $\widetilde{\vect{t}}$ is defined in $\{0\} \cup [\underline{c},1]$, hence, any $\widetilde{\vect{t}}$ belongs to $\mathbb{H}_u^N$.\\
Now, comparing $\vect{t}_m$ and the observation vector $\widetilde{\vect{t}} \in \mathbb{H}_u^N$ gathered in correspondence to the occurrence of the $m$-th event, two types of discrepancies can emerge. One consists on a bias between the two vectors: this happens when no misdetections nor misclassifications occur and it is only due to the confidence of the classification algorithm (i.e., $c_n$ if $v_n \in \mathcal{V}_m$). The other one arises when the vector $\widetilde{\vect{t}}$ is projected from $\vect{t}_m$ to another vertex \mbox{$\vect{t}_{m^\prime} \in \mathbb{V}\mathbb{H}_u^{N,0} \setminus \vect{t}_m$}. This is a consequence of a misdetection or a misclassification: from~\eqref{eq:h_i_1_reformulate}, the former implies that non-zero entries in $\vect{t}_m$ become null in $\widetilde{\vect{t}}$, while the latter has the opposite effect. If $\vect{t}_{m^\prime}$ is an active vertex, the observation vector is the result of an artifact; if $\vect{t}_{m^\prime}$ is a non-active vertex, the {\emph{importance}} of $e_m$ and $e_{m^\prime}$ is altered (see Section~\ref{subsec:GMM_interpretation}).\\
Figure~\ref{fig:cube} provides a visual representation of the given geometrical interpretation of the SM and of the network observations, with $N=3$ and $M=4$. 
\end{rem}

\section{SM Identification via GMM}
\label{sec:identification}

This section aims at presenting a solving procedure for the SM identification problem. In particular, the outlined strategy takes advantage of the interpretation of the given observations model as GMM.

\subsection{Observations Model GMM Interpretation}\label{subsec:GMM_interpretation}

As highlighted in Remark~\ref{rem:geometric_interpretation}, any observation vector $\widetilde{\vect{t}}$, $m \in [1,M]$, constitutes a point on the hypercube $\mathbb{H}_u^{N}$. In particular, when dealing with real-world application, it holds that $\underline{c} \approx 1$, and then $\widetilde{\vect{t}}$ belongs to the neighborhood of one of the hypercube vertices.

\begin{example}
To clarify this fact, let us consider a VSN composed of $N=2$ cameras. In this case the probability~\eqref{eq:network_likelihood_iid} results to be
{\setlength{\mathindent}{0.2cm} \small
\begin{align}
\label{eq:example_N_2}
    	p(\widetilde{\vect{t}} \vert e_m) 
    	 \!=\!& \sum_{k_1=0}^1  \sum_{k_2=0}^1 \left((1-\alpha_{m,1})\mathcal{U}(\widetilde{t}_1 ; \underline{c},1)\right)^{k_1} \! \left( \alpha_{m,1}\delta(\widetilde{t}_1)\right)^{1-k_1}  \nonumber\\[-0.1cm]
    	& \qquad \quad  \left((1-\alpha_{m,2})\mathcal{U}(\widetilde{t}_2 ; \underline{c},1)\right)^{k_2}\left( \alpha_{m,2}\delta(\widetilde{t}_2)\right)^{1-k_2}  \\
    	 = & \alpha_{m,1}\alpha_{m,2}\delta(\widetilde{t}_1)\delta(\widetilde{t}_2) + \nonumber\\
    	 & \; + \alpha_{m,1}(1-\alpha_{m,2})\delta(\widetilde{t}_1)\mathcal{U}(\widetilde{t}_2;\underline{c},1)+ \nonumber\\
    	& \; + (1-\alpha_{m,1})\alpha_{m,2}\delta(\widetilde{t}_2)\mathcal{U}(\widetilde{t}_1 \vert \underline{c},1)+ \nonumber \\
    	& \; + (1-\alpha_{m,1})(1-\alpha_{m,2})\mathcal{U}(\widetilde{t}_1 ; \underline{c},1)\mathcal{U}(\widetilde{t}_2 ; \underline{c},1). 
\end{align}}
Based on~\eqref{eq:example_N_2}, if event $e_m$ occurs in the monitored environment, the observation vector $\widetilde{\vect{t}}$ is equal to $[0 \; 0]^\top$, namely it coincides with the origin of a unitary square, with probability $\alpha_{m,1}\alpha_{m,2}$. Alternatively, it is close to the vertex  $[0 \; 1]^\top$ with probability $\alpha_{m,1}(1-\alpha_{m,2})$ and to the vertex  $[1 \; 0]^\top$ with probability $(1-\alpha_{m,1})\alpha_{m,2}$, and the proximity is determined by a uniform distribution over the non-zero component. Finally, $\widetilde{\vect{t}}$ approximates the vertex $[ 1 \; 1]^\top$ with probability $(1-\alpha_{m,1})(1-\alpha_{m,2})$: the proximity depends on a bi-variate uniform distribution over the two components of the observation vector, so that it holds $\underline{c} \leq \widetilde{t}_{1},\widetilde{t}_{2} \leq 1$.
To clarify, if $\vect{t}_m = [ 0 \; 1]^\top$, then $\alpha_{m,1}(1-\alpha_{m,2})$ represents the probability of correct observation, while the other terms in~\eqref{eq:example_N_2} represent observation errors and artifacts.\\
Notably, accounting for the normal approximation of a uniform distribution~\cite{krivanek2003fast} and applying the total probability theorem, the generative probability distribution of any network observation $\widetilde{\vect{t}}$ can be approximated by the following 
% Furthermore, it is reasonable to assume independence between camera measurements; therefore it is possible to approximate the stimulation likelihood
% $p(\widetilde{\vect{t}}_m|e_m)$ with a multivariate \mbox{$N$-dimensional} Gaussian distribution, namely
% \begin{equation}\label{eq:gaussian_approx}
% 	\begin{split}
% 		& \widetilde{\vect{t}}_m := \widetilde{\vect{t}}|e_m \sim \mathcal{N}(\widetilde{\vect{t}}|\bm{\mu}_m, \bm{\Sigma}_m) \\
% 		& \bm{\mu}_m \approx \vect{t}_m \\
% 		& \bm{\Sigma}_m = diag(\sigma_1,\dots,\sigma_N).
% 	\end{split}
% \end{equation}
% Then, by applying the Total Probability Theorem on the random vector $\widetilde{\vect{t}}$, the following
GMM, which depends on the parameter set $\widetilde{\bm{\phi}} =\{ \phi_m \} _{m=1}^4$, as described in Table~\ref{tab:GMM_parameters_N_2}, namely
\begin{equation}
\label{eq:GMM_interpretation_N_2}
	\begin{split}
		 p_{\widetilde{\bm{\phi}}}(\widetilde{\vect{t}}) \approx \sum_{m=1}^{4} \widetilde{\pi}_m \mathcal{N}(\widetilde{\vect{t}} \vert \widetilde{\bm{\mu}}_m, \widetilde{\bm{\Sigma}}_m).
	\end{split}
\end{equation} 
\end{example}

\begin{table}[t!]
	\centering
	\caption{Parameters of the GMM~\eqref{eq:GMM_interpretation_N_2}}
	\label{tab:GMM_parameters_N_2}
	 \resizebox{.9\columnwidth}{!}{
	\begin{tabular}{|c| ccc|}
		\cline{2-4}
		\multicolumn{1}{c|}{} & $\widetilde{\pi}_m$  & $\widetilde{\bm{\mu}}_m$      & $\widetilde{\bm{\Sigma}}_m$\\
		\hline
		$\phi_1$ & $\sum_{\ell=1}^M p(e_{\ell})\alpha_{\ell,1}\alpha_{\ell,2}$ & $\mathbf{0}_2$ &
        $\mathbf{0}_{2\times 2}$
		\\
		$\phi_2$ & $\sum_{\ell=1}^M  p(e_{\ell})\alpha_{\ell,1}(1-\alpha_{\ell,2})$ & 
		$\frac{\underline{c}+1}{2}\begin{bmatrix}
		    0 & 1
		\end{bmatrix}^\top$
		&
     	$\begin{bmatrix}
		    0 & \frac{(1-\underline{c})^2}{12}
		\end{bmatrix}\mathbf{I}_{2\times 2}$
		\\
		$\phi_3$ & $\sum_{\ell=1}^M p(e_\ell)(1-\alpha_{\ell,1})\alpha_{\ell,2}$ & 
		$\frac{\underline{c}+1}{2}\begin{bmatrix}
		    1 & 0
		\end{bmatrix}^\top$
		&
		$\begin{bmatrix}
		    \frac{(1-\underline{c})^2}{12} & 0
		\end{bmatrix}\mathbf{I}_{2\times 2}$
		\\
		$\phi_4$ & $\sum_{\ell=1}^M p(e_\ell)(1-\alpha_{\ell,1})(1-\alpha_{\ell,2})$ &
		$\frac{\underline{c}+1}{2}\begin{bmatrix}
		    1 & 1
		\end{bmatrix}^\top$
		&
		$\frac{(1-\underline{c})^2}{12}\mathbf{I}_{2\times 2}$
		\\
		\hline
	\end{tabular}
	}
\end{table} 

The reasoning carried out in the given example can be generalized to the case of a VSN made up of $N \geq 2$ cameras. In particular, it holds that
\begin{equation}\label{eq:GMM_measurements}
		p_{\widetilde{\bm{\phi}}}({\widetilde{\vect{t}}}) \approx \sum_{m=1}^{2^N} \widetilde{\pi}_m \mathcal{N}(\widetilde{\vect{t}} \vert  \widetilde{\bm{\mu}}_m, \widetilde{\bm{\Sigma}}_m).
\end{equation} 
Note that any observation turns out to be approximately distributed according to a $N$-dimensional GMM with a number of components equal to the number of vertices of the $N$-dimensional hypercube, i.e., to $2^N$. In detail, it holds that {$\widetilde{\bm{\mu}}_m \approx \vect{t}_m$, for all $m \in [1,M]$}, given that $\underline{c} \approx 1$. {Moreover, from the generalization of Table~\ref{tab:GMM_parameters_N_2}, it follows that any weight $\widetilde{\pi}_m$, representing the importance of the $m$-th GMM component, can be expressed as
{\setlength{\mathindent}{0.2cm} \begin{align}
    \widetilde{\pi}_m 
    & = \sum_{\ell=1}^M p(e_\ell)\prod_{n=1}^N \left(\alpha_{\ell,n}-2\alpha_{\ell,n}q_{m,n} + q_{m,n}\right)  \\
    %     & = \left(\sum_{\ell:\vect{t}_\ell = \vect{v}_{m}} p(e_\ell)\prod_{n=1}^N \left(\alpha_{\ell,n}-2\alpha_{\ell,n}q_{m,n} + q_{m,n}\right)\right)+ \nonumber \\
    % &\qquad  + \left(\sum_{\ell:\vect{t}_\ell \neq \vect{v}_{m}} p(e_\ell)\prod_{n=1}^N \left(\alpha_{\ell,n}-2\alpha_{\ell,n}q_{m,n} + q_{m,n}\right)\right)\\
    & = {\widetilde{\pi}_m^\prime}+{\widetilde{\pi}_m^{\prime\prime}}, \end{align}}
with $q_{m,n}$ being the $n$-th entry of the $m$-th vector $\vect{q}_{m} \in \mathbb{V}\mathbb{H}_u^{N}$ and
{\setlength{\mathindent}{0.2cm} \small \begin{align}
    \widetilde{\pi}_m^\prime  & = \begin{cases}
     p(e_m)\prod\limits_{n=1}^N \left(\alpha_{m,n}-2\alpha_{m,n}q_{m,n} + q_{m,n}\right) & {\text{if} \;} e_m \in \mathcal{E}_a  \\ 0  & \text{otherwise}
    \end{cases} \label{eq:importance_explicit_2} \\
        \widetilde{\pi}_m^{\prime\prime} & = \sum_{\ell: e_\ell \in \mathcal{E}_a \setminus \{e_m\}} p(e_\ell)\prod_{n=1}^N \left(\alpha_{\ell,n}-2\alpha_{\ell,n}q_{m,n} + q_{m,n}\right). \label{eq:importance_explicit_1}
\end{align}}
The term $\widetilde{\pi}_m^\prime$ in~\eqref{eq:importance_explicit_2}  represents the a-priori probability of the $m$-th event, revised in the light of the cameras detection and recognition capabilities, namely combined with the probabilities $p_D$ and $p_C$ encoded in the quantities $\lbrace \alpha_{m,n}\rbrace_{n=1}^N$. On the other hand, the term  $\widetilde{\pi}_m^{\prime\prime}$ in~\eqref{eq:importance_explicit_1} represents an alteration of the $m$-th vertex importance, caused by the (incorrect) classification of any event $e_{m^\prime}$, $m^\prime \neq m$, as $e_m$ according to the $m^\prime$-th event occurrence probability and cameras detection and recognition capabilities. Mathematically, this results from the sum of $M-1$ or $M$ components, given that $\vect{q}_m$ corresponds to an active vertex or not.}
In this sense, in the ideal case wherein $p_C=p_D =1$, it holds from~\eqref{eq:h_i_1_reformulate}
\begin{equation}
    \alpha_{m,n} =
    1-t_{m,n}
\end{equation}
then, accounting for~\eqref{eq:importance_explicit_1}, it follows that
\begin{align}
    & \widetilde{\pi}_m^\prime =
    \begin{cases}
        p(e_m) =\pi_m, & m \in [1,M] \\
        0, & m \in [M+1,2^N]
    \end{cases}\\
    & \widetilde{\pi}_m^{\prime\prime} = 0
\end{align}
Thus, it is possible to conclude that
\begin{equation}\label{eq:GMM_interpretation}
		p_{\bm{\phi}}(\widetilde{\vect{t}}) \approx \sum_{m=1}^{M} p(e_m) \mathcal{N}(\widetilde{\vect{t}} \vert \vect{t}_m, \bm{\Sigma}_m),
\end{equation} 
namely, the VSN stimulation likelihood given in Section~\ref{subsec:observation} can be interpreted as a GMM, parameterized by {$\bm{\phi} = \left\lbrace p(e_m), \; \vect{t}_m,\; \bm{\Sigma}_m \right\rbrace_{m=1}^M \in \Phi$} and whose number of components corresponds to the number $M$ of active events.

Given these premises, one can realize that it is possible to address the SM identification problem as a canonical GMM fitting task, and, thus exploiting the standard statistical methods to estimate the parameter set $\bm{\phi}$~\cite{tzikas2008variational,distributedEM_finite}.  %distributedEM,,VariationalGMM
% Then, from \eqref{eq:gaussian_approx}
% \begin{equation}\label{eq:T_sim_mu}
	% \vect{T} \approx \begin{bmatrix}
		% \bm{\mu}_1^\top & \dots & \bm{\mu}_M^\top  
		% \end{bmatrix}^\top.
	% \end{equation}
 In view of the foregoing, the following proposition shows how the fact that the SM is associated to the not necessary injective function~\eqref{eq:selective_function} (Remark~\ref{R1}) affects the solution of Problem~\ref{prob:stimulation_model_identification}.
\begin{prop}\label{prop:identifiability}
	If the SM $\tau^\prime(\cdot)$ is not injective, then the stimulation matrix \eqref{eq:stimulation_matrix} is not \textit{globally identifiable}.
	%the identification problem is not feasible in the global identifiability sense \cite{rothenberg1971identification}.
	
\end{prop}
\begin{proof}
	As defined in~\cite{rothenberg1971identification}, a family of probability densities $\{p_{\bm{\phi}}(\cdot), \bm{\phi} \in \Phi\}$ is globally identifiable if 
	\begin{equation}\label{eq:identifiability}
		p_{\bm{\phi}_1}(\cdot) = p_{\bm{\phi}_2}(\cdot) \Rightarrow \bm{\phi}_1 = \bm{\phi}_2, \; \forall \bm{\phi}_1,\bm{\phi}_2 \in \Phi,
	\end{equation}
	while the fact that $\tau^\prime(\cdot)$ is not injective means that
	\begin{equation}
		\exists m^\prime,m^{\prime\prime} \in [1,M] \text{ s.t. } m^\prime \neq m^{\prime\prime} \wedge \vect{t}_{m^\prime} = \vect{t}_{m^{\prime\prime}}.
	\end{equation}
	Then, from~\eqref{eq:GMM_interpretation}, it follows that
	{\setlength{\mathindent}{0.2cm} \small
	\begin{align}\label{eq:non_identifiability_derivation}
	p_{\bm{\phi}_1}({\widetilde{\vect{t}}}) 
			& \approx \sum_{m=1}^M \pi_m \mathcal{N}(\widetilde{\vect{t}} \vert \vect{t}_m, \bm{\Sigma}_m) \nonumber \\
			& \approx \sum_{m=1}^{M-2} \pi_m \mathcal{N}(\widetilde{\vect{t}} \vert \bm{\vect{t}}_m, \bm{\Sigma}) + \pi_{m^\prime} \mathcal{N}(\widetilde{\vect{t}} \vert \bm{\vect{t}}_{m^\prime}, \bm{\Sigma}) + \pi_{m^{\prime\prime}} \mathcal{N}(\widetilde{\vect{t}} \vert \bm{\vect{t}}_{m^{\prime\prime}}, \bm{\Sigma})  \nonumber \\
			& = \sum_{m=1}^{M-2} \pi_m \mathcal{N}(\widetilde{\vect{t}} \vert \bm{\vect{t}}_m, \bm{\Sigma}) + (\pi_{m^\prime} + \pi_{m^{\prime\prime}}) \mathcal{N}(\widetilde{\vect{t}} \vert \bm{\vect{t}}_{m^\prime}, \bm{\Sigma})  \nonumber \\
			& = \sum_{m=1}^{M-1} \pi_m^{\prime} \mathcal{N}(\widetilde{\vect{t}} \vert \bm{\vect{t}}_m, \bm{\Sigma})  \nonumber \\
			& \approx \sum_{m=1}^{M-1} \pi_m^{\prime} \mathcal{N}(\widetilde{\vect{t}} \vert \vect{t}_m, \bm{\Sigma}_m) = p_{\bm{\phi}_2}({\widetilde{\vect{t}}}),
		\end{align}}
	where $\vect{\Sigma}_{m^{\prime}} = \vect{\Sigma}_{m^{\prime\prime}}$, from the hypothesis $\vect{t}_{m^{\prime}}=\vect{t}_{m^{\prime\prime}}$. From \eqref{eq:non_identifiability_derivation} it follows that the underlying GMM distribution $p_{\bm{\phi}_1}(\widetilde{\vect{t}})$ is indistinguishable with respect to $p_{\bm{\phi}_2}(\widetilde{\vect{t}})$, which is a GMM with $M-1$ components. 
\end{proof}
From Proposition~\ref{prop:identifiability}, one can conclude that when the function~\eqref{eq:selective_function} is not injective), namely when the stimulation matrix $\vect{T}$ in~\eqref{eq:stimulation_matrix} is not full-row rank, Problem~\ref{prob:stimulation_model_identification} is ill-posed, given that $\bm{\phi}$ is not identifiable. In other words, any observation vector $\widetilde{\vect{t}}$ can not provide complete information about the parameter set $\bm{\phi}$. For this reason, the following assumption is stated. Note that this requires to design the network deployment with balanced degree of redundancy~\cite{VarottoSelection}.%,hoblos2000fault}. %,staroswiecki2004sensor 

\begin{assumption}
    The function $\tau(\cdot)$ in~\eqref{eq:selective_function} introduced in Definition~\ref{def:stimulation_model} is supposed to be injective.
\end{assumption}

Typically, learning problems suffer from the so-called \textit{curse of dimensionality}, especially in correspondence to large scale networks~\cite{poggio2018theory}. Also the discussed SM identification is affected by this issue. %Indeed, in this framework, such a phenomenon arises as soon as the number of cameras $N$ becomes large. 
In this case, an AE is employed to reduce the data dimensionality, enabling the implementation of the deep embedded clustering over a lower-dimensional feature space while preserving the local data structure properties~\cite{goodfellow2016deep}. In this manner, Problem~\ref{prob:stimulation_model_identification} boils down to a particular GMM inference problem, namely the estimation of GMM parameters retrieved from the network observations dataset~\eqref{eq:dataset_network}.

\subsection{Identification
Procedure}\label{subsec:pipeline}

\begin{figure*}[t!]
	\centering
	\subfloat[\label{fig:pipeline_training}]{
		\includegraphics[trim={0.35cm 0 0.6cm 0},clip,width=0.25\textwidth]{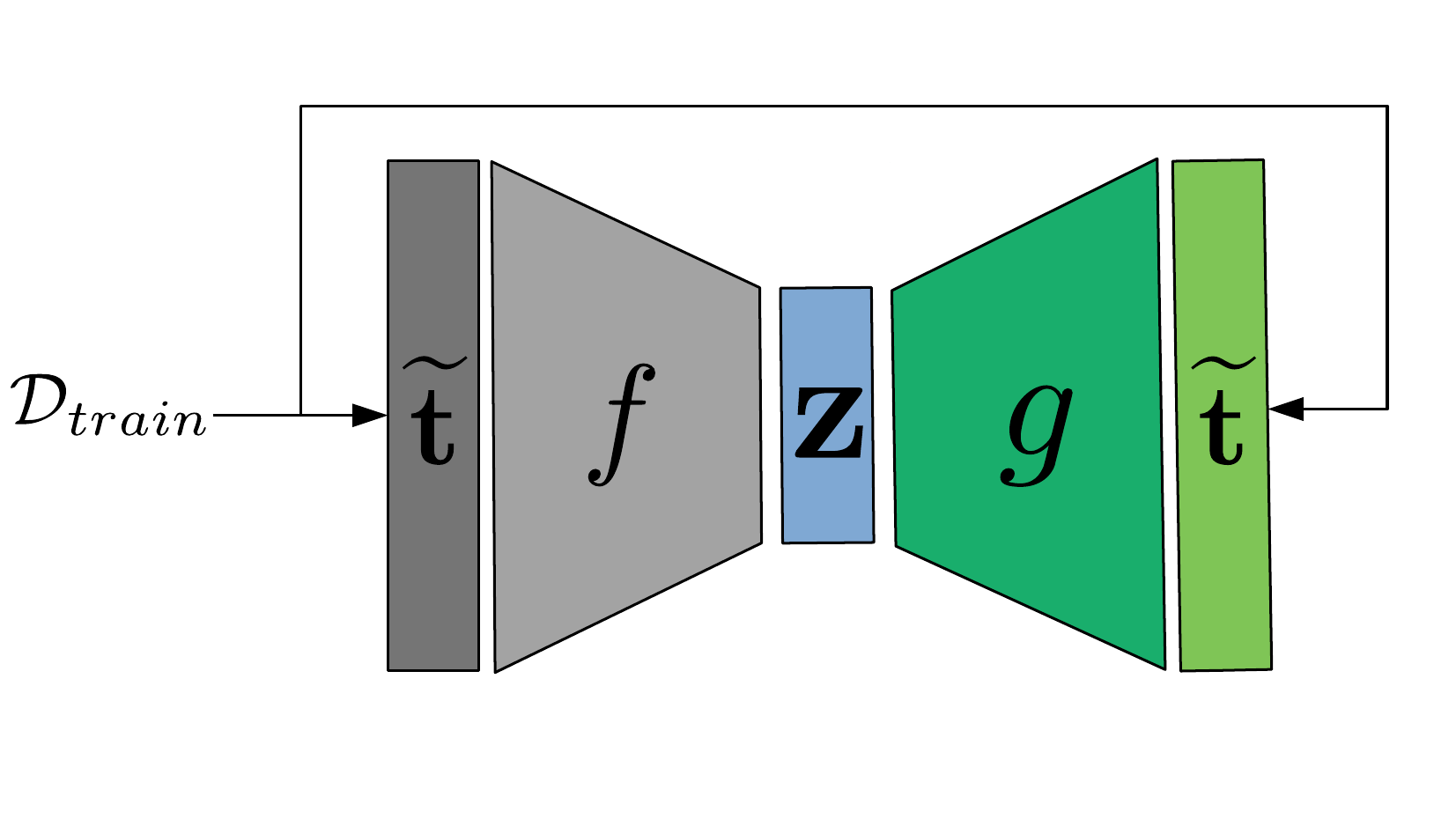}}%
	~ 
    \hspace{0.5cm}
	\subfloat[\label{fig:pipeline_clustering}]{
		\includegraphics[trim={0.5cm 0 0.6cm 0},clip,width=0.25\textwidth]{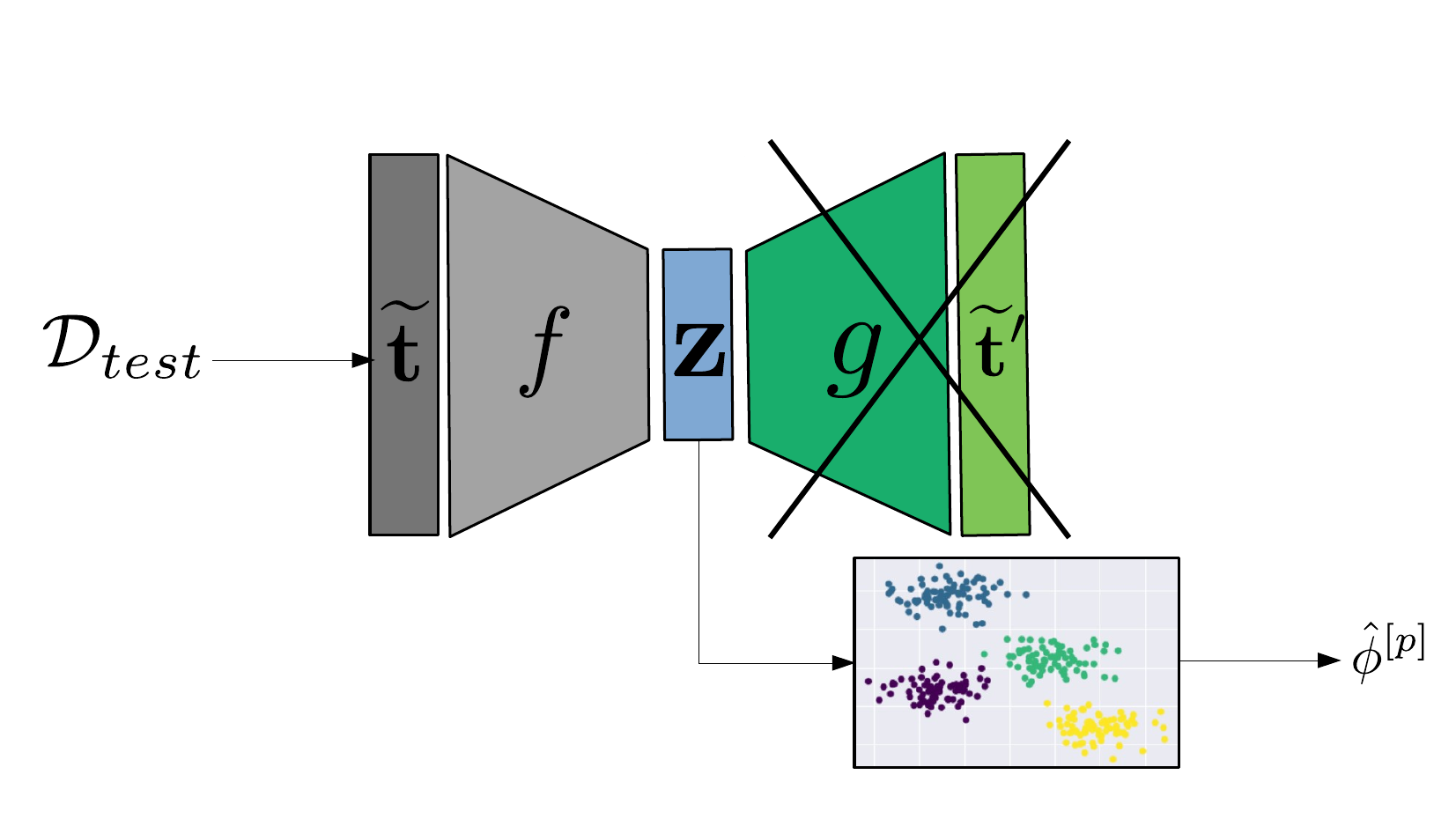}}
	~ 
	\hspace{-0.7cm}
	\subfloat[\label{fig:pipeline_decoder}]{
		\includegraphics[trim={0.5cm 0 1cm 0},clip,width=0.25\textwidth]{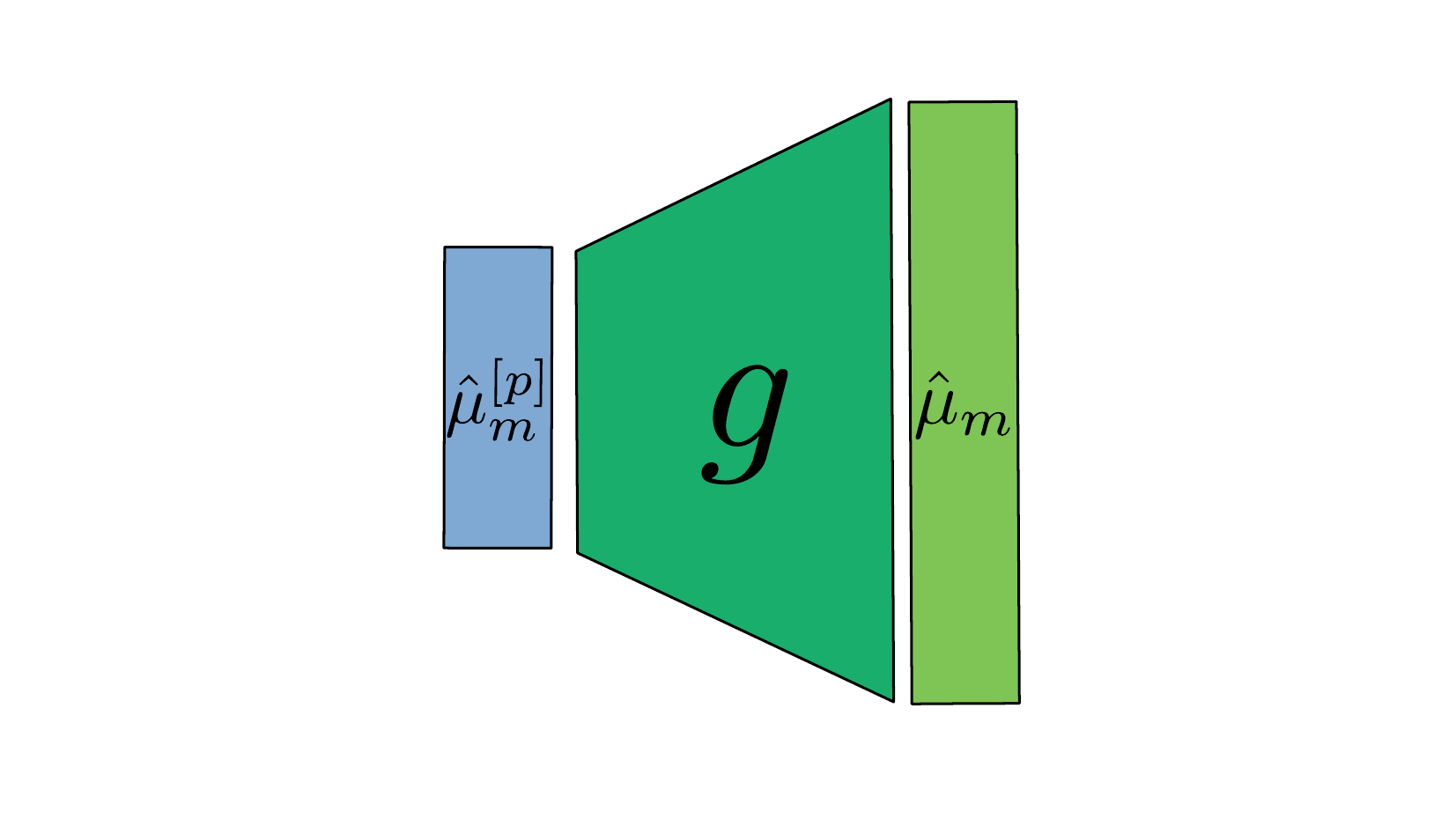}}%
	~ 
	\hspace{-0.5cm}
	\subfloat[\label{fig:pipeline_vertices}]{
		\includegraphics[trim={2.5cm 0 1cm 0},clip,width=0.25\textwidth]{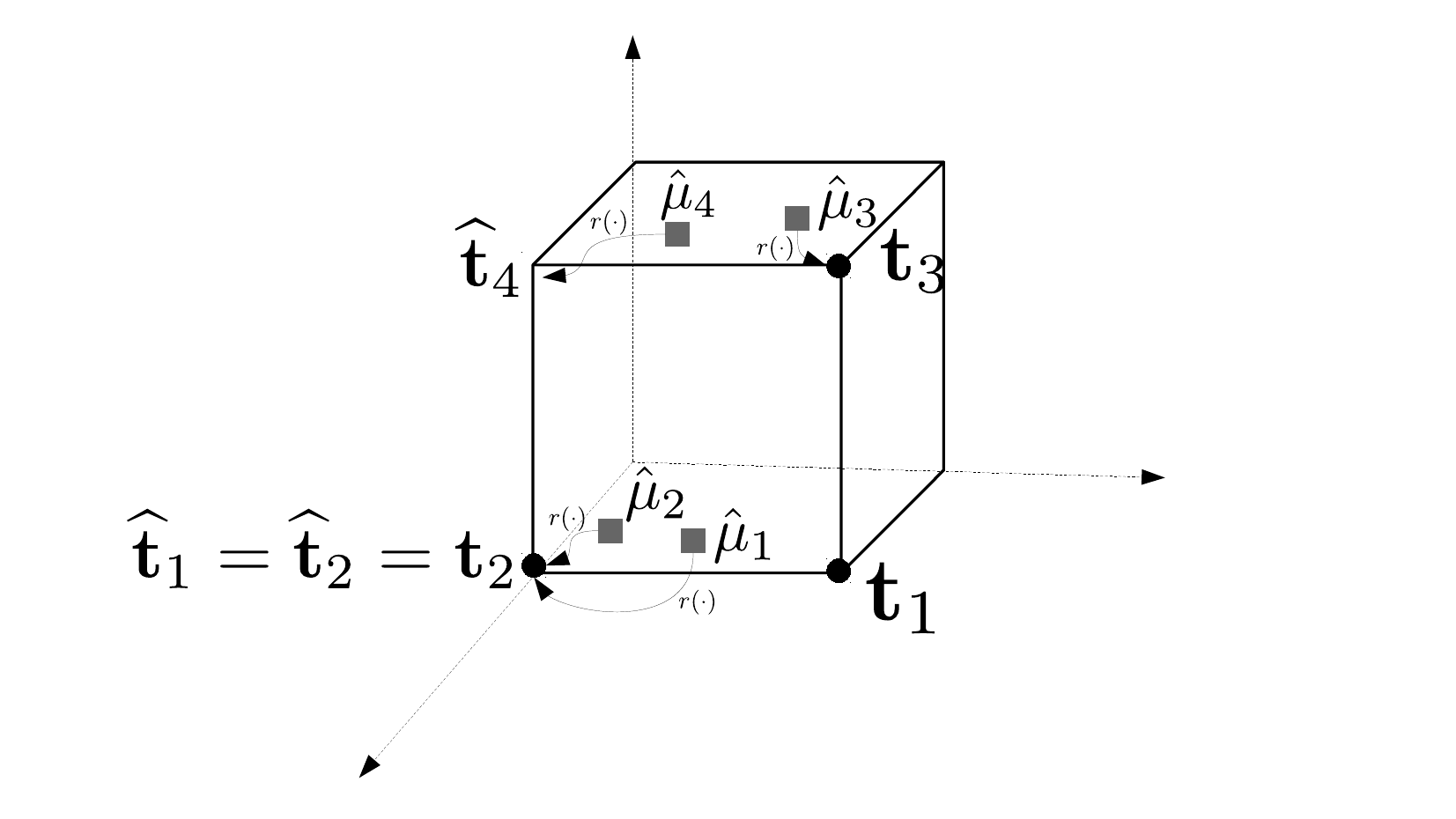}}
	\caption{
		\footnotesize 
		Deep embedded features clustering pipeline: (a) AE training; (b) deep embedded features clustering; (c) exploitation of the decoder to project the GMM centroids defined in a $p$-dimensional space onto $\mathbb{H}_u^N$; (d) rounding process of the GMM centroids from $\mathbb{H}_u^N$ to $\mathbb{V}\mathbb{H}_u^N$ through $r(\cdot)$ - in this case, it is $N=3$, $M=3$, $\hat{M}=4$, and $\widehat{M}_{eff}=3$ ($\widehat{\bm{\mu}}_1$ and $\widehat{\bm{\mu}}_2$ are both projected onto the same vertex $\vect{t}_2$).
	}
	\label{fig:pipeline}
\end{figure*}
% trim={<left> <lower> <right> <upper>}

Hereafter, the outlined procedure aiming at solving Problem~\ref{prob:stimulation_model_identification} is described. This is based on the given observation that the SM identification problem can be addressed as a canonical GMM fitting task, leveraging~\eqref{eq:GMM_interpretation}. In this sense, the curse of dimensionality emerging in correspondence to high values of the number $N$  of devices composing the VSN are faced through the introduction of an AE. 

It is worth to recall that the dimensionality reduction techniques are fundamental to decrease the amount of undesired artifacts generated by the employed observations/measurements~\cite{dimensionalityReduction_tutorial,dimensionalityReduction_review}, and in the proposed solution, the AE is specifically used to reduce the dimension of the GMM to learn since the number of fitting GMM components is typically greater than the number of events $M$.  On the other hand, it is also  to be noted that, in a compact space, the soft-partition clustering can be well combined with the deep embedding~\cite{li2021softpartitionclustering}. As a consequence, an AE can be generally adopted a-fortiori to also discard the low-prior GMM components.  

In light of these facts, Figure~\ref{fig:pipeline} illustrates the overall proposed SM identification strategy. This entails the four principal steps described in the following.

%\LV{Evidenziare che la riduzione della dimensionalità è fondamentale anche per ridurre gli artefatti creati dalle misure (il numero di componenti GMM è maggiore di $M$!). In uno spazio compatto, il soft-clustering dovrebbe scartare le componeti a bassa prior}

%To conclude, the recent VSNs literature also accounts for the issues deriving from the increasing number of the devices composing the networks. Several dimensionality reduction techniques are, indeed, well-known~\cite{dimensionalityReduction_tutorial,dimensionalityReduction_review}. %
%{Among them, the solutions based on the AEs (see, e.g.~\cite{wang2016auto,DL_tutorial,charte2018practical}) % 
%are characterized by well-stated theoretical function approximation properties and demonstrated learning capabilities since these are based the deep neural networks~\cite{bengio2013representation}

\begin{itemize}[leftmargin = 0.1cm]
	\item[] \textbf{AE training -}The observations dataset $\mathcal{D}$ introduced in~\eqref{eq:dataset_network} is generated according to the stochastic characterization described in Section~\ref{subsec:observation}. This is successively split into the \textit{training} and \textit{testing} subsets as follows
	\begin{equation}
		\begin{split}
			& \mathcal{D}_{train} = \left\lbrace \widetilde{\vect{t}}^d \right\rbrace_{d=1}^{D_{train}}    \\
			& \mathcal{D}_{test} = \mathcal{D} \setminus \mathcal{D}_{train} =  \left\lbrace \widetilde{\vect{t}}^d \right\rbrace_{d=D_{train}+1}^{D}.
		\end{split}
	\end{equation}
	The subset $\mathcal{D}_{train}$ is then used to train an AE adopting the unsupervised learning  paradigm (Figure~\ref{fig:pipeline_training}). 
	
	\item[] \textbf{Data encoding \& GMM learning -} The subset $\mathcal{D}_{test}$  is provided as input to the encoder part of the AE, which aims at compressing its $N$-dimensional data points into the $p$-dimensional embedded features $\left\lbrace \vect{z}^d \right\rbrace_{d=D_{train}+1}^{D}$  (Figure~\ref{fig:pipeline_clustering}). These features are then used to figure out the following GMM 	\begin{equation}\label{eq:GMM_estimated_smallDim}
		p_{\widehat{\bm{\phi}}^{[p]}}(\vect{z}) = \sum_{m=1}^{\widehat{M}} \widehat{\pi}_m^{[p]} \mathcal{N}(\mathbf{z} \vert \widehat{\bm{\mu}}_m^{[p]}, \widehat{\bm{\Sigma}}_m^{[p]} ) ,
	\end{equation} 
	where the notation $(\cdot)^{[p]}$ is used to stress the fact that the probability distribution~\eqref{eq:GMM_estimated_smallDim} is defined in $\mathbb{R}^p$ (contrarily to the one in \eqref{eq:GMM_interpretation} defined in $\mathbb{H}_u^N$), while the notation $\hat{(\cdot)}$ identifies the estimated quantities. These last correspond to the GMM parameters, and include also the number $\widehat{M}$ of the GMM components.

	\item[] \textbf{Data decoding -} Since it is defined in a different (reduced) space, the probability distribution~\eqref{eq:GMM_estimated_smallDim} does not properly describe the SM. The decoder part of the AE is thus used to project each $p$-dimensional vector $\widehat{\bm{\mu}}_m^{[p]}$ onto the (original) $N$-dimensional space $\mathbb{H}_u^N$, obtaining $\widehat{\bm{\mu}}_m \in \mathbb{H}_u^N$ (Figure~\ref{fig:pipeline_decoder}). The decoder does not modify neither the number of GMM components nor their corresponding weights, hence it theoretically holds that
	\begin{align}
			& \widehat{M} = M  \label{eq:GMM_approx_1}\\
			& \widehat{\pi}_m = \pi_m, \; m \in [1,\widehat{M}] \\
			& \vect{t}_m \approx 
			\widehat{\bm{\mu}}_m, \; m \in [1,\widehat{M}]\label{eq:GMM_approx_3}
		\end{align}
	where $M$ and $\left\lbrace \pi_m \right\rbrace_{m=1}^M$ are the parameters characterizing~\eqref{eq:GMM_interpretation}. 
	Note that, in this work, a two-stage clustering is proposed, as in~\cite{tian2014learning}; however, in the literature there are attempts to jointly accomplish feature learning and clustering in a single-stage training process~\cite{caron2018deep}. 
	%CAE,yang2016joint

	\item[] \textbf{Rounding operation-} {The approximation $\vect{t}_m \approx 
	\widehat{\bm{\mu}}_m$ in~\eqref{eq:GMM_approx_3} is due to the fact that {$\bm{\mu}_m \in \mathbb{H}_u^N$} and {$\vect{t}_m \in \mathbb{V}\mathbb{H}_u^{N,0}$}. The two quantities coincide} by  projecting any $\bm{\mu}_m$ onto $\mathbb{V}\mathbb{H}_u^{N,0}$ (Figure~\ref{fig:pipeline_vertices}). This is equal to rounding each entry of the vector $\bm{\mu}_m$ to the closer binary number, i.e., to either $0$ or $1$. The rounding operation can be formalized through the introduction of the following map
	\begin{equation}\label{eq:rounding_map}
			r: \;
			\mathbb{H}_u^N \rightarrow \mathbb{V}\mathbb{H}_u^{N,0}, \quad \widehat{\bm{\mu}}_m \mapsto \widehat{\vect{t}}_m.
	\end{equation}
	It is to be noted that the map $r(\cdot)$ is not injective in general, as a consequence, it follows that
	\begin{equation}\label{eq:M_eff}
		{| \Ima r\left(\mathbb{H}_u^N\right) |:=\widehat{M}_{eff} \leq \widehat{M}}
	\end{equation}
	where $\widehat{M}_{eff}$ corresponds to the effective estimated number of GMM components. 
\end{itemize}

In the light of~\eqref{eq:M_eff}, it is suitable to distinguish between the \textit{estimated stimulation matrix} $\widehat{\vect{T}} \in \{0,1\}^{\widehat{M} \times N}$, defined as
\begin{equation}
		\widehat{\vect{T}} = 
		\begin{bmatrix}
			\widehat{\vect{t}}_1 & \dots & \widehat{\vect{t}}_{\widehat{M}}
		\end{bmatrix}^\top,
	\end{equation}
 and the \textit{effective estimate of the stimulation matrix} $\widehat{\vect{T}}_{eff} \in  \{0,1\}^{\widehat{M}_{eff} \times N}$. The latter depends on the former as follows 
	\begin{equation}\label{eq:Teff}
		\widehat{\vect{T}}_{eff} =   \widehat{\vect{E}} \widehat{\vect{T}},
	\end{equation}
where $\widehat{\vect{E}} \in \{0,1\}^{\widehat{M}_{eff} \times \widehat{M}}$ acts by removing the duplicated rows in $\widehat{\vect{T}}$. This matrix is, indeed, constructed by setting  $[\widehat{\vect{E}}]_{ii} = 1$, $[\widehat{\vect{E}}]_{ij} = 0$, $i\neq j$, when the $i$-th row in $\widehat{\vect{T}}$ is redundant (i.e., it is linearly dependent on other rows) and $[\widehat{\vect{E}}]_{ij} = 1$,  $i\neq j$, otherwise. 	Similarly, the \textit{effective GMM estimated weights} are defined as
	\begin{equation}\label{eq:effective_weights}
		\widehat{\pi}_{m,eff} = \sum_{k \in [1,\widehat{M}] : \; \widetilde{\vect{t}}_k = \widetilde{\vect{t}}_m } \widehat{\pi}_k, \quad m\in[1,\widehat{M}_{eff}].
	\end{equation}
	Any $\widehat{\pi}_{m,eff} \in [0,1]$ combines the weights of all the GMM components projected by $r(\cdot)$ onto the same vertex of $\mathbb{H}_u^N$.

\subsection{Performance Assessment}

The performance of the proposed AE-based identification procedure can be assessed by evaluating the \emph{reconstruction error}. This is defined as the sum of two (positive) terms, namely  
{\setlength{\mathindent}{0.2cm}
\begin{align}\label{eq:rec_err}
		& e_r = e_r^{{at}} + e_r^{{md}}  \\
		& e_r^{{at}} = \frac{1}{\overline{M}-M} \sum_{k=1}^{\widehat{M}_{eff}} \mathbbm{1} \left( {\widehat{\vect{t}}_k} \not\in \Ima \tau^\prime([1,M]) \right) \widehat{\pi}_{k,eff}  \label{eq:rec_err_at} \\
		& e_r^{{md}} = \frac{1}{M} \sum_{m=1}^{M} \mathbbm{1} \left( \vect{t}_m \not\in {\Ima r\left(\mathbb{H}_u^N\right)} \right) \pi_m  \label{eq:rec_err_md}
\end{align}}
where $\mathbbm{1} (\cdot)$ is an indicator function, which is equal to $1$ in correspondence to a true condition and $0$ otherwise. 
The term~\eqref{eq:rec_err_at} accounts for the artifacts  generated by the identification procedure, and  for their weight; in other words, $e_r^{{at}}$ takes into account the GMM centroids that are projected onto non-active vertices of $\mathbb{H}_u^N$. The term~\eqref{eq:rec_err_md}, instead, accounts for the active vertices not detected by the identification procedure, and their weights. Thus, one can conclude that the reconstruction error $e_r$ accounts for the importance of both spurious and miss-detected events. 

The following proposition shows that the error~\eqref{eq:rec_err} constitutes a valid performance index to determine the effectiveness of the proposed procedure to solve Problem~\ref{prob:stimulation_model_identification}.
% Recalling that the objective of this work is to estimate the stimulation matrix $\vect{T}$, the Proposition below proves that $e_r$ is a valid performance index to evaluate the identification capability of the proposed solution. 

\begin{prop}\label{prop: perfect_reconstruction}
% 	Assuming that the estimated GMM components in \eqref{eq:GMM_approx} are proper, meaning that $\widehat{\pi}_{k,eff} \neq 0$, $\forall k= 1 \ldots \widehat{M}_{eff}$, the reconstruction error in \eqref{eq:rec_err} is null if and only if
% 	\begin{equation}\label{eq:ernullrelation}
% 		\widehat{\vect{T}}_{eff} = \vect{P} \vect{T},
% 	\end{equation}
% 	where $\vect{P} \in \{0,1\}^{M \times M}$ is a permutation matrix.
Assuming that the effective GMM estimated weights~\eqref{eq:effective_weights} are proper, meaning that $\widehat{\pi}_{k,eff} \neq 0$, for all $k \in [1,\widehat{M}_{eff}]$, the reconstruction error $e_r$ in~\eqref{eq:rec_err} is null if and only if $\widehat{\vect{T}}_{eff} = \vect{P} \vect{T}$ with  $\vect{P} \in \{0,1\}^{M \times M}$ being a permutation matrix.
\end{prop}
\begin{proof}
	{Trivially, since $e_{r}^{at} \geq 0$, $e_{r}^{md} \geq 0$, it holds that $e_{r} = 0$ if and only if both $e_{r}^{at} = 0$ and $e_{r}^{md} = 0$. In addition, exploiting the assumption $\widehat{\pi}_{k,eff} \neq 0$, $\forall k \in [1, \widehat{M}_{eff}]$, it also holds that $e_{r}^{at} = 0$ if and only if $\widehat{\vect{t}}_{k} \in \Ima \tau, ~ \forall k \in [1, \widehat{M}_{eff}]$, and   $e_{r}^{md} = 0 $ if and only if $ \vect{t}_{m} \in \Ima r\left(\mathbb{H}_u^N\right), ~ \forall m \in[1,M]$.
	Hence, it follows that $e_{r}$ is null if and only if \begin{equation}\label{eq:ernullcond}
		\begin{cases}
			\hat{\vect{t}}_{k} \in \Ima \tau^\prime([1,M]), & \forall k \in [1, \widehat{M}_{eff}],\\
			\vect{t}_{m} \in \Ima r\left(\mathbb{H}_u^N\right), & \forall m \in[1,M].
		\end{cases}
	\end{equation}
	The two conditions in~\eqref{eq:ernullcond} are equivalent to the following inclusions
	\begin{equation}
		\begin{cases}
			\Ima \vect{\widehat{T}}_{eff}^{\top} \subseteq  \Ima \vect{T}^{\top}, \\
			\Ima \vect{T}^{\top}  \subseteq \Ima \vect{\widehat{T}}_{eff}^{\top},
		\end{cases}
	\end{equation}
        % 	The first condition in~\eqref{eq:ernullcond} is equivalent to the inclusion $\Ima \vect{\widehat{T}}_{eff}^{\top} \subseteq  \Ima \vect{T}^{\top} $; while the second condition is equivalent to the inclusion $ \Ima \vect{T}^{\top}  \subseteq \Ima \vect{\widehat{T}}_{eff}^{\top} $.
        	leading to the requirement $\Ima \vect{\widehat{T}}_{eff}^{\top} = \Ima \vect{T}^{\top}$. This implies that the equivalence $\widehat{M}_{eff} = M$ is a necessary condition to ensure $e_{r} = 0$. As a consequence, $e_{r}$ is null if and only if, {for all $m \in [1,M]$}, the $m$-th row of the matrix $\vect{\widehat{T}}_{eff}$ is equal to the $m$-th row of the product matrix $\vect{P}\vect{T}$, {given some permutation induced by $\vect{P}$}. Matrix $\vect{P}$ is uniquely determined, since each row both in $\vect{T}$ and in $\vect{\widehat{T}}_{eff}$ is a $N$-dimensional transpose column vector linearly independent from the other rows. % (see also construction \eqref{eq:Teff}). 
        	This observation concludes the proof. %leads to the thesis; indeed, $e_r = 0$ if and only if \eqref{eq:ernullrelation} is verified.
        	}
\end{proof}
{Remarkably, Proposition \ref{prop: perfect_reconstruction} shows that  zeroing the reconstruction error~\eqref{eq:rec_err} corresponds to solve Problem~\ref{prob:stimulation_model_identification}, since it is a necessary and sufficient to ensure that 
%is zero if and only if 
the effective estimated stimulation matrix $\vect{\widehat{T}}_{eff}$ is equal to the stimulation matrix $\vect{T}$ up to a row permutation. Furthermore, from the proof of Proposition~\ref{prop: perfect_reconstruction}, the following corollary ensues.}
\begin{corol}\label{corol:M_eff_er}
	If $\widehat{M}_{eff} \neq M$, then $e_r > 0$. 
\end{corol}

Note that, in Figure~\ref{fig:pipeline_vertices}, it holds that $e_r^{{at}} = \widehat{\pi}_4$, since $\widetilde{\vect{t}}_4 \not\in \Ima \tau^\prime([1,M])$, and $e_r^{{md}} =\pi_1$, since $\vect{t}_1 \not\in  \Ima r\left(\mathbb{H}_u^N\right)$.

%%%%%%%%%%%%%%%%%%%%%%%%%%%%%%%%%%

% NUMERICAL RESULTS
\section{Numerical Results}
\label{sec:numerical_results}

This section reports the results of the experimental campaign carried out on a fully synthetic dataset to investigate the performance of the proposed SM identification procedure\footnote{The code can be found at \url{github.com/luca-varotto/VSN-with-AE}}. The outcome of a single illustrative test is first discussed for the purpose of providing an insight on the outlined strategy. Then, the results of a Monte Carlo (MC) experiment are presented aiming at strengthening the solution validity.

In detail, hereafter, two versions of the method described in Section~\ref{subsec:pipeline} are taken into account distinguishing between GMM+AE-based and V-GMM+AE-based identification. The former learns the {$p$-dimensional} GMM~\eqref{eq:GMM_estimated_smallDim} through the EM algorithm, while the latter exploits the VB inference. Table~\ref{tab:AE_parameters} reports the parameters related to the AE structure and its training process, performed using the Keras deep learning library~\cite{AE_keras}. %It is worth noting that the feature space $\mathcal{F}$ belongs to a plane (i.e., $p=2$),while the use of a sigmoid at the last decoder layer forces the re-projections onto $\mathbb{H}_u^N$.
Such two versions of the outlined solution for Problem~\ref{prob:stimulation_model_identification} are compared with the following alternative SM identification methods.
% , \ToDo{constituting a more basic and more complex approach, respectively}.
 
\begin{itemize}
\item{\textit{(V-)GMM}:} the SM identification is faced by applying the canonical GMM fitting methods on the original space $\mathbb{H}_u^N$. In detail, the \mbox{$N$-dimensional} GMM~\eqref{eq:GMM_interpretation} is directly learnt on the input data, either via EM algorithm (GMM) or through VB inference (V-GMM). 
The comparison between (V-)GMM and (V-)GMM+AE is meant to prove that, to obtain acceptable identification performance, it is necessary to apply a dimensionality reduction scheme.

\item{\textit{(V-)GMM+DNN}:}  the AEs mitigate the effects of the curse of dimensionality, but they apply a lossy transformation on the input data, with possible detrimental effects on the overall identification pipeline; therefore, the \mbox{(V-)GMM+DNN}-based solution class is taken into account. In this case, the undercomplete AE is replaced by an overcomplete one\footnote{Here it is adopted the same taxonomy proposed in \cite{charte2018practical}.} (later on generically referred to as Deep Neural Network - DNN). In particular, the encoder of the considered DNN has two layers, each composed by $N$ nodes, so that a lossless transformation is applied from $\mathbb{H}_u^N$ to $\mathbb{H}_u^N$ itself. The training process exploits the parameters in {Table~\ref{tab:AE_parameters}}. \\
The comparison between (V-)GMM+DNN $~$ and $~$ (V-)\\GMM+AE is meant to prove that, with regard to the GMM inference process, the detrimental effects of the curse of dimensionality (present in the DNN version) are more critical than the lossy transformation (present in AE version).

%\item{\ToDo{Test other dim. reduction techniques, e.g. \mbox{(V-)GMM+PCA}}}.
\end{itemize}

\subsection{Single Illustrative Test}\label{subsec:run}

% Figure~\ref{fig:VGMM_AE_clusters} shows the clustering of the deep embedded features in $\mathbb{R}^2$ obtained via V-GMM.
% \begin{figure}[t!]
	%     \centering
	%         \includegraphics[width=0.3\textwidth]{./images/VGMM_AE_clusters.png}
	%         \caption{\footnotesize
		% Clustering of embedded features via V-GMM.}
	%         \label{fig:VGMM_AE_clusters}
	% \end{figure}%

With the aim of exemplifying the working principles of the proposed SM identification solution, a VSN composed of $N=15$ cameras is taken into account for a single test. The network observations are generated according to the protocol described in Section~\ref{subsec:observation} and based on the setup parameters reported in Table~\ref{tab:setup_parameters}. The (real) stimulation matrix $\vect{T}$ is, instead, randomly selected.

\begin{table}[t!]
	\centering
	\caption{
		Main parameters regarding the encoder/decoder structure and the training process.}
	\label{tab:AE_parameters}
	\begin{tabular}{|l|c|}
		\hline
		number of layers & $5$\\ 
		\rowcolor{lightgray}
		nodes per layer & $ N-12-8-4-2 \quad (p=2)$ \\ 
		number of epochs & $15$\\ 
		\rowcolor{lightgray}
		batch size & $30$\\ 
		loss function & MSE \\ 
		\rowcolor{lightgray}
		optimizer & Adadelta\\ 
		activation function & 
		ReLU (last layer decoder: sigmoid)\\
		\hline
	\end{tabular}
\end{table}

\begin{table}[th!]
	\centering
	\caption{Single Illustrative run. Principal setup parameters.}
	\label{tab:setup_parameters}
	\begin{tabular}{|l|c|}
		\hline
%		number of cameras & $N=15$ \\ 
		number of potential events & $\overline{M} = 20$\\
				\rowcolor{lightgray}
		number of active events & $M=3$ \\ 
		events generator pdf & $e_m \sim \mathcal{U}(1,M), m \in [1,M]$ \\ 
				\rowcolor{lightgray}
		observations set cardinality & $D=10^4$ \\ 
        probability of detection & $p_D = 0.8$ \\ 
				\rowcolor{lightgray}
		probability of classification & $p_C = 0.99$ \\
		minimum confidence value & $\underline{c}=0.7$ \\
    		\rowcolor{lightgray}
		confidence value & 
		$c_n \sim \mathcal{U}(\underline{c},1), \; n \in [1,N]$\\
		\hline
	\end{tabular}
\end{table} 

As described in Section~\ref{subsec:pipeline}, the SM identification performance can be assessed via the reconstruction error~\eqref{eq:rec_err}. In addition,  the following quantities are hereafter considered.
\begin{itemize}
	% \item Clustering accuracy $acc$: as described in Section~\ref{subsec:pipeline}, one of the core tasks of the overall pipeline is GMM learning, which can be seen as a soft clustering problem (Section~\ref{subsec:AE}). The accuracy is computed applying the Hungarian algorithm \cite{kuhn1955hungarian} to the confusion matrix \cite{roc_analysis}. 
	\item \textit{estimated number of active events} $\widehat{M}_{eff}$:  it results to be interesting to compare $\widehat{M}_{eff}$ with $\widehat{M}$. Indeed, when these variables are different, their comparison allows to understand if the identification procedure is affected by overfitting ($\widehat{M}_{eff} > M$) or underfitting ($\widehat{M}_{eff} < M$). 
	Moreover,  this reveals the importance of the rounding operation through the map~\eqref{eq:rounding_map}.  
	\item \textit{Kullback-Leibler (KL) divergence} $D_{KL}$: %\GMi{of what?}
	even though the focus of this work is on the identification of the active events stimulation actions (via the estimation of the matrix $\vect{T}$), it is also worth studying the capability of retrieving the events a-priori distribution. 
	To this aim, it is possible to compute the KL divergence %\cite{D_KL} 
	between $\left\lbrace \pi_m \right\rbrace_{m=1}^M$ and $\left\lbrace \widehat{\pi}_{m,eff} \right\rbrace_{m=1}^{\widehat{M}_{eff}}$. 
	However, accounting for the fact that the two distributions might not have the same support (i.e., whenever $e_r \neq 0$), it results to be more convenient to consider the following modified version of the KL divergence:
	\begin{equation}
	\label{eq:KL_divergence}
		D_{KL} = \frac{1}{M_\tau} \sum_{m \text{ : } \vect{t}_m \not\in \Ima r\left(\mathbb{H}_u^N\right)} \pi_m \log \frac{\pi_m }{ \widehat{\pi}_{m,eff} },
	\end{equation}
	where $M_\tau \leq M$ is the number of \textit{detected} active events (note that if $M_\tau=0$, $D_{KL}$ is not defined).
% 	\todo[inline]{$M$ or $M_\tau$?}
\end{itemize}
%\vspace{0.2cm}

Given these premises, Table~\ref{tab:performance_uniform} reveals that all the considered SM identification methods experience a certain level of overfitting, since $\widehat{M}_{eff} > M$ in all the cases. Nonetheless, this fact is particularly evident for the (V-)GMM solution class, where $\widehat{M}_{eff}=\overline{M}=20$; on the other side, the (V-)GMM+AE procedures turn out to be more accurate ($\widehat{M}_{eff} \approx M$), together with GMM+DNN. However, the main difference between the two AE-based solutions and GMM+DNN resides in the value of the reconstruction error, which is remarkably smaller when adopting 
% the proposed (V-)GMM+AE approach, 
the former approaches,
meaning that all active events are detected and the few introduced artifacts have negligible importance weight. 

\begin{table}[t!]
	\centering
	\caption{
		Single illustrative test. Performance indexes of the six algorithms under comparison: uniform events prior case. \emph{Best figures are highlighted in bold.}}
	\label{tab:performance_uniform}
	\begin{tabular}{|l cccc|}
		\hline
		$ $ & $\widehat{M}$  & $\widehat{M}_{eff}$      & $e_r$ & $D_{KL}$\\
		\hline
		GMM      & $20$      & $20$      & $0.161$ & $0.487$\\
		\rowcolor{lightgray}
		V-GMM    & $20$      & $20$      & $0.039$ & $0.386$ \\ \hline
				GMM+AE   & $6$      & $\bm 4$      & $\bm{0.003}$  & $\bm{0.017}$\\
		\rowcolor{lightgray}
		V-GMM+AE & $14$      & $6$      & $0.013$  & $0.081$\\ \hline
		GMM+DNN  & $4$      & $\bm 4$      & $0.392$  & / \\
		\rowcolor{lightgray}
		V-GMM+DNN  & $20$      & $16$      & $0.038$  & $0.561$ \\
		\hline
	\end{tabular}
\end{table} 

\begin{table}[t!]
	\centering
	\caption{Single illustrative test. Performance indices of the six algorithms under comparison: non-uniform events prior case. \emph{Best figures are highlighted in bold.}}
	\label{tab:performance_nonUniform}
	\begin{tabular}{|c cccc|}
		\hline
		$ $ & $\hat{M}$  & $\hat{M}_{eff}$      & $e_r$ & $D_{KL}$\\
		\hline
				GMM      & $20$      & $20$      & $0.041$ & $0.394$\\
		\rowcolor{lightgray}
		V-GMM    & $20$      & $20$      & $0.045$ & $0.508$ \\ \hline
		GMM+AE   & $7$      & $\bm 3$      & $\bm{0.0}$  & ${3.4\times10^{-4}}$\\
		\rowcolor{lightgray}
		V-GMM+AE & $18$      & $\bm{3}$      & $\bm{0.0}$  & $\bm{1.6\times10^{-4}}$\\ \hline
		GMM+DNN  & $12$      & $8$      & $0.026$  & $0.216$ \\
		\rowcolor{lightgray}
		V-GMM+DNN  & $20$      & $15$      & $0.014$  & $0.092$ \\
		\hline
	\end{tabular}
\end{table}

\begin{figure*}[t!]
	\centering
	\subfloat[\label{fig:weights}]{
		\includegraphics[width=0.48\textwidth]{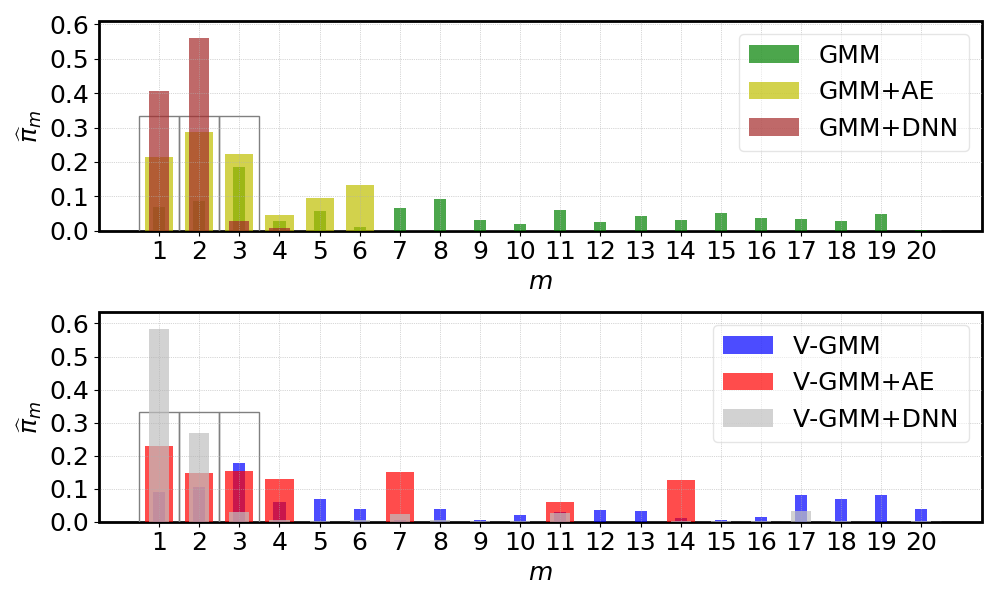}}%
	~ 
	\subfloat[\label{fig:weights_eff}]{
		\includegraphics[width=0.48\textwidth]{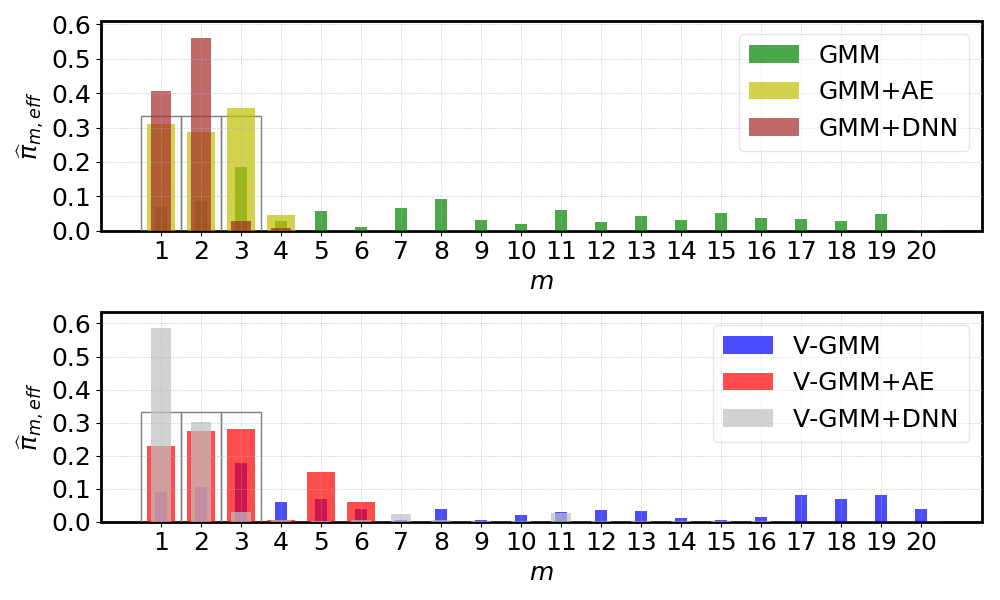}}
	\caption{
		Single illustrative test given $N=15$. Uniform events prior (unfilled boxes) and estimates (filled boxes): \textbf{(a)} before rounding process; \textbf{(b)} after rounding process.}
	\label{fig:pdf_reconstruction_uniform}
\end{figure*}

\begin{figure*}[t!]
	\centering
	\subfloat[\label{fig:weights_nonUNiform}]{
		\includegraphics[width=0.49\textwidth]{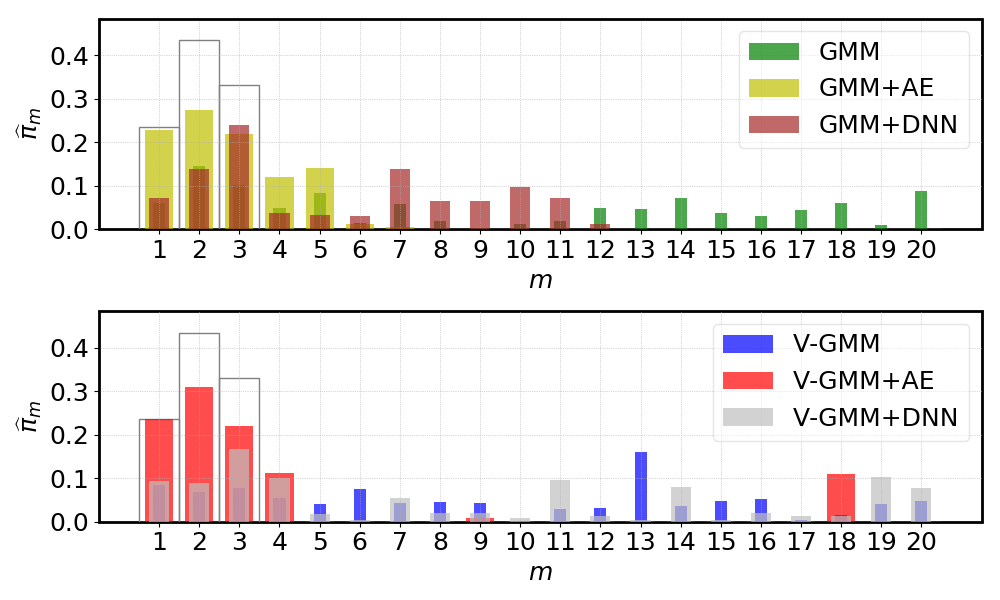}}%
	~ 
	\subfloat[\label{fig:weights_eff_nonUNiform}]{
		\includegraphics[width=0.49\textwidth]{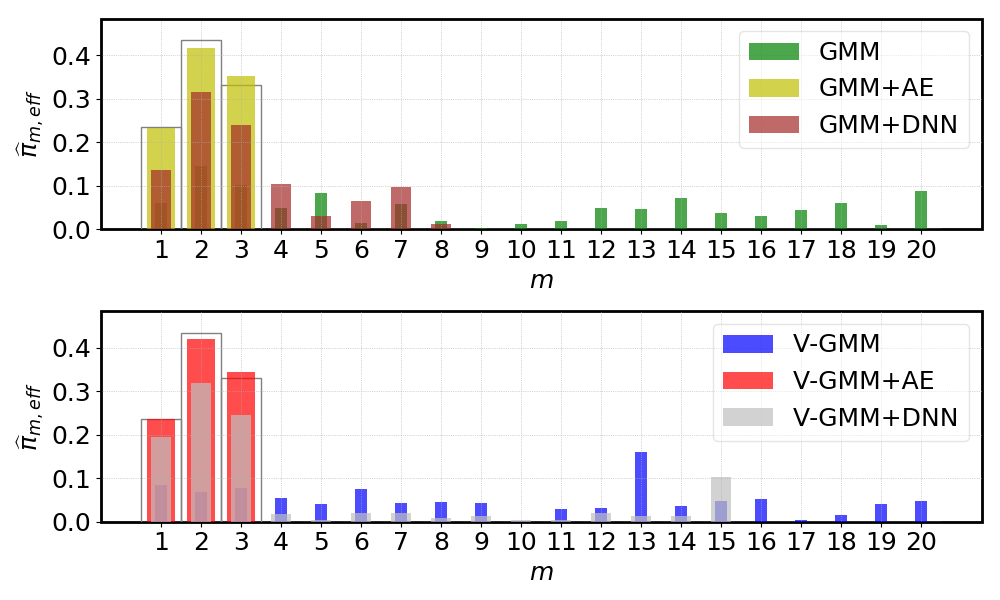}}
	\caption{
		Single illustrative test given $N=15$. Non-uniform events prior (unfilled boxes) and estimates (filled boxes): \textbf{(a)} before rounding process; \textbf{(b)} after rounding process.}
	\label{fig:pdf_reconstruction_nonUniform}
\end{figure*}

The results reported in Table~\ref{tab:performance_uniform} are graphically depicted in Figure~\ref{fig:pdf_reconstruction_uniform}\footnote{In general, the results obtained with the EM-based approach are characterized by lighter overfitting than the variational counterpart. 
It is important to recall that V-GMM chooses automatically the optimal number of components, while GMM requires an extensive cross-validation phase.}. 
In detail, Figure~\ref{fig:weights} represents the estimated events prior before the rounding process, while Figure~\ref{fig:weights_eff} shows the final results after the rounding process. In both figures, the underlying uniform event priors are depicted with unfilled boxes. 
Hence, the effect of the rounding process \eqref{eq:rounding_map} can be evaluated by comparing Figure~\ref{fig:weights} with Figure~\ref{fig:weights_eff}: in the former, a large number of spurious components characterizes the estimated event distributions for all the considered methods; in the latter, this issue is partially solved, due to the aggregation of multiple centroids on a single vertex, according to~\eqref{eq:effective_weights}.
From Figure~\ref{fig:weights_eff} one can notice that the GMM+AE solution provides the most accurate prior estimate (confirmed also by the smallest value of $D_{KL}$ in Table~\ref{tab:performance_uniform}), while V-GMM+AE yields a slightly worse result in terms of overfitting (even though the reconstruction capability is still acceptable if compared with all the other methods, as suggested by Table~\ref{tab:performance_uniform}).
Moreover, the already mentioned overfitting effect in (V-)GMM, as well as in V-GMM+DNN, is evident from the large number of spurious components that appear in the final estimates. Conversely, the GMM+DNN method experiences a milder overfitting effect; nonetheless, the reconstruction process is completely ruinous, as validated also by the non-defined value of $D_{KL}$ and the high value of $e_r$ reported in Table~\ref{tab:performance_uniform}.  

Interestingly, the (V-)GMM+AE approaches outperform the other identification solutions also in the case of non-uniform events prior distribution, as shown by Table~\ref{tab:performance_nonUniform} and Figure~\ref{fig:pdf_reconstruction_nonUniform}. 
Overall, similar considerations to the uniform case can be done; for instance, the beneficial effect of the rounding procedure is evident also in this scenario (see Figure~\ref{fig:weights_nonUNiform}). In addition, in this case both GMM+AE and V-GMM+AE solutions entails the reconstruction error zeroing, even though the events distribution estimate is not without errors, as can be seen in the last column of Table~\ref{tab:performance_nonUniform} and in Figure~\ref{fig:weights_eff_nonUNiform}. On the other side, GMM+DNN still experiences poor reconstruction capabilities, denoted by the value $e_r = 0.026$, together with a higher overfitting effect with respect to the uniform case ($\widehat{M}_{eff} = 8$ in Table~\ref{tab:performance_nonUniform} against $\widehat{M}_{eff} = 4$ in Table~\ref{tab:performance_uniform}).

Finally, it is worth mentioning the fact that, in the ideal case of $p_D=1$ and $\underline{c}=0.9$, all the considered strategies manage to perfectly identify the SM (i.e., $e_r=0$); moreover, all of them provide an accurate estimate of the events a-priori distribution ($D_{KL} < 10^{-3}$). This is due by the fact that $p_D=1$ prevents the generation of artifacts, while $\underline{c}=0.9$ ensures that each observation $\widetilde{\vect{t}}^{d}$ is close to the corresponding active vertex $\vect{t}_m$, whenever the $m$-th event occurs in the monitored environment; hence, the collected data are intrinsically aggregated on $M$ compact clusters and the GMM learning becomes easy even at high dimensions.

\subsection{MC experiment}\label{subsec:MC}

The discussed single illustrative test is useful to clarify the working principle of the proposed  SM identification procedure and its benefits with respect to other possible solutions. However, in highly dynamic scenarios characterized by elevated variability in terms of parameters, a single test is not fully representative of the method performance. In particular, one can observe that for a $N$-devices VSN and a certain set of $M$ events, it is possible to have $\prod_{i=0}^{M}\left(2^N - i\right)$ different stimulation matrices. Furthermore, in Section~\ref{subsec:observation} the nature of the network observations is assumed to be intrinsically stochastic, and also the adopted GMM learning techniques are based on optimization procedures which do not guarantee the global optimality of their solution.

For these reasons, hereafter, the results of an MC experiment are discussed, focusing on all the variable contributes in the SM identification process. More precisely, the single illustrative test described in Section~\ref{subsec:run} is repeated $50$ times. In correspondence to each run, a stimulation matrix is randomly generated and a synthetic dataset is determined on the basis of the computed $\vect{T}$ and  according to Table~\ref{tab:setup_parameters}.

The performance are evaluated by accounting for:
\begin{itemize}
	\item the histograms of $\widehat{M}$ and $\widehat{M}_{eff}$ over the $50$ MC runs, to quantify possible overfitting effects;
	\item  the Empirical Cumulative Distribution Function\\ (ECDF) of the reconstruction error~\eqref{eq:rec_err}  and of the KL divergence~\eqref{eq:KL_divergence} over the $50$ MC runs, defined as
\begin{equation}\label{eq:ECDF}
% 		\hat{F}^{(50)}_{\xi}(a) = \frac{1}{50} \sum_{i=1}^{50}\mathbbm{1}_{\xi \leq a}
	\hat{F}^{(50)}_{\bullet}(a) = \frac{1}{50} \sum_{i=1}^{50} f_\bullet(a)
	\end{equation}
	with $\bullet$ standing for $e_r$ or $D_{KL}$ and $f_\bullet(a)$  denoting the indicator function, assuming value $1$ only when the corresponding performance index is smaller than a certain threshold $a$, and $0$ otherwise.
\end{itemize}
\vspace{0.2cm}

\begin{figure}[t]
	\centering
	\subfloat[\label{fig:ECDF_er}]{
		\includegraphics[width=0.4\textwidth]{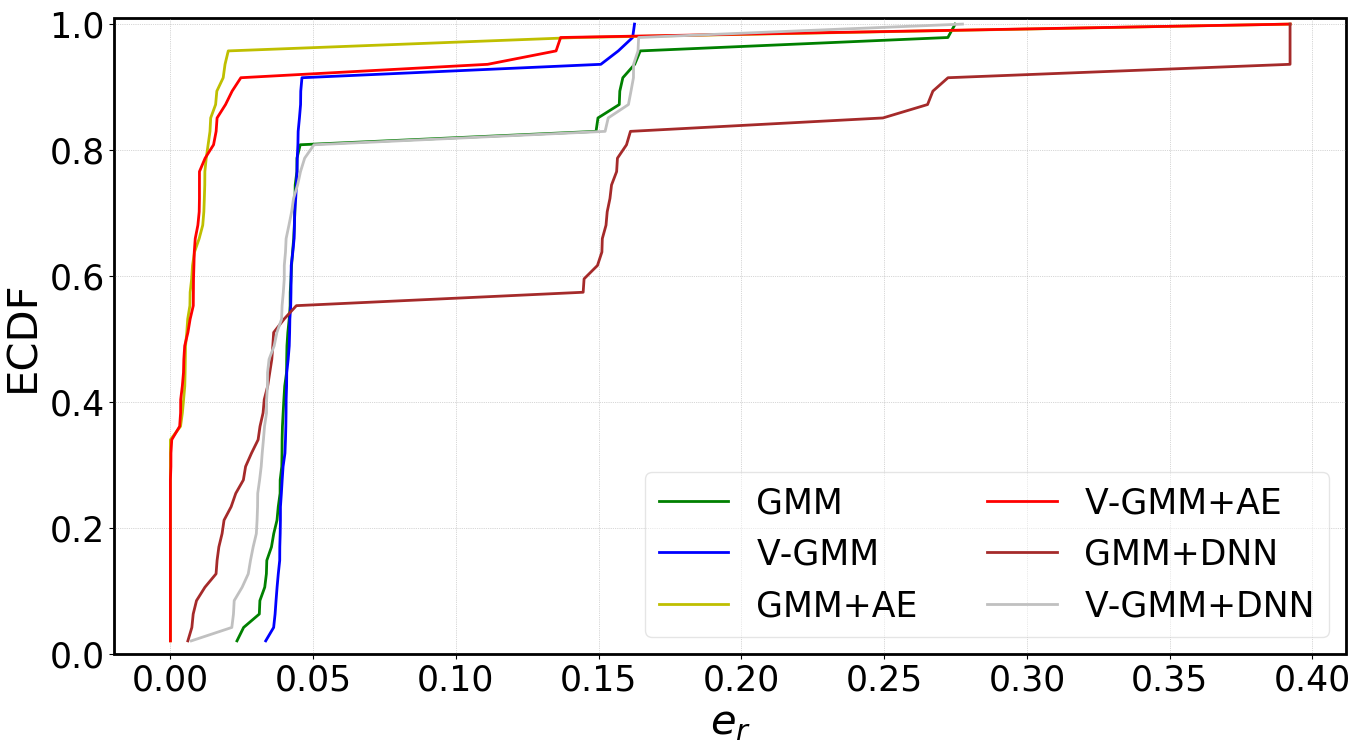}}\\
	\subfloat[\label{fig:ECDF_KL}]{
		\includegraphics[width=0.4\textwidth]{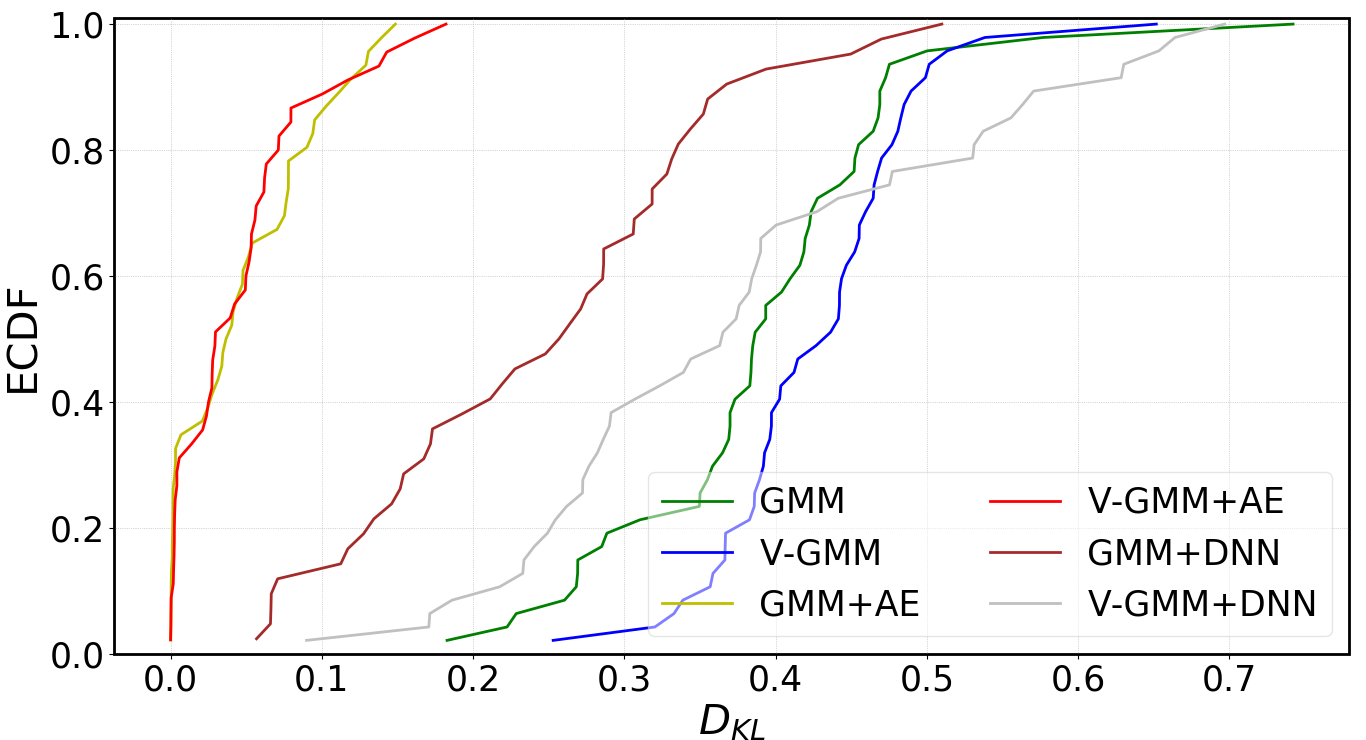}}
	\caption{MC experiment given $N=15$. ECDF of $e_r$ \textbf{(a)} and $D_{KL}$ \textbf{(b)} over the $50$ MC runs.}
	\label{fig:ECDF}
\end{figure}

From Figure~\ref{fig:ECDF_er}, it is possible to note that the weight of the spurious/non-detected components tends to be much smaller when adopting the outlined \mbox{(V-)GMM+AE} approach with respect to the other identification methods. Furthermore, this solution outperforms both \mbox{(V-)GMM} and \mbox{(V-)GMM+DNN} identification methods in terms of capability of retrieving the a-priori distribution of the active events. This fact is justified by the results in Figure~\ref{fig:ECDF_KL}: the $D_{KL}$ index is never larger than $0.2$ when accounting for the \mbox{(V-)GMM+AE} approach, while the same value is overcome more than $60\%$ of times by all the other strategies. 

\begin{figure*}[t]
	\centering
	\subfloat[\label{fig:weights_MC}]{
		\includegraphics[width=0.49\textwidth]{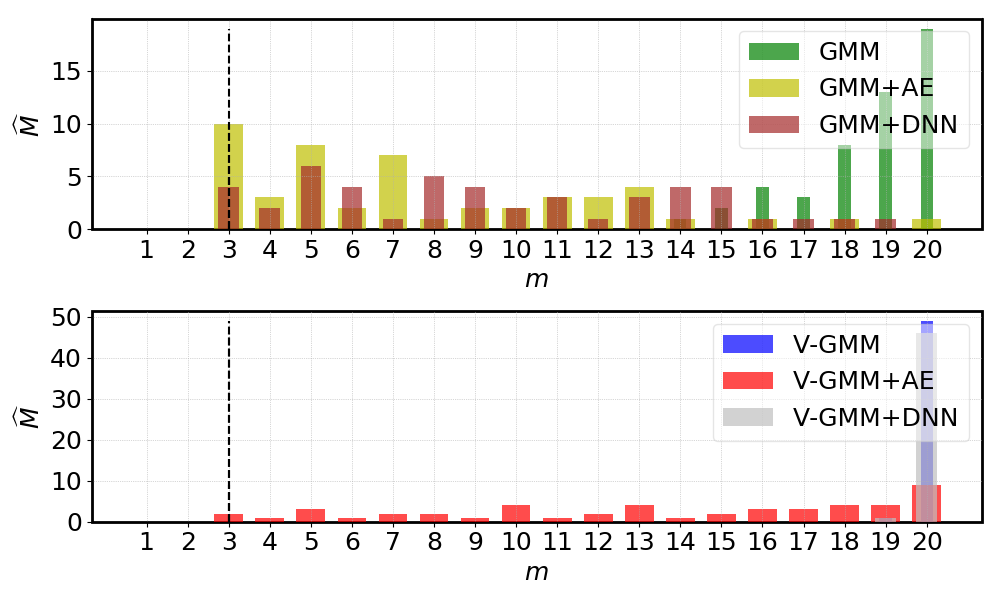}}%
	~ 
	\subfloat[\label{fig:weights_eff_MC}]{
		\includegraphics[width=0.49\textwidth]{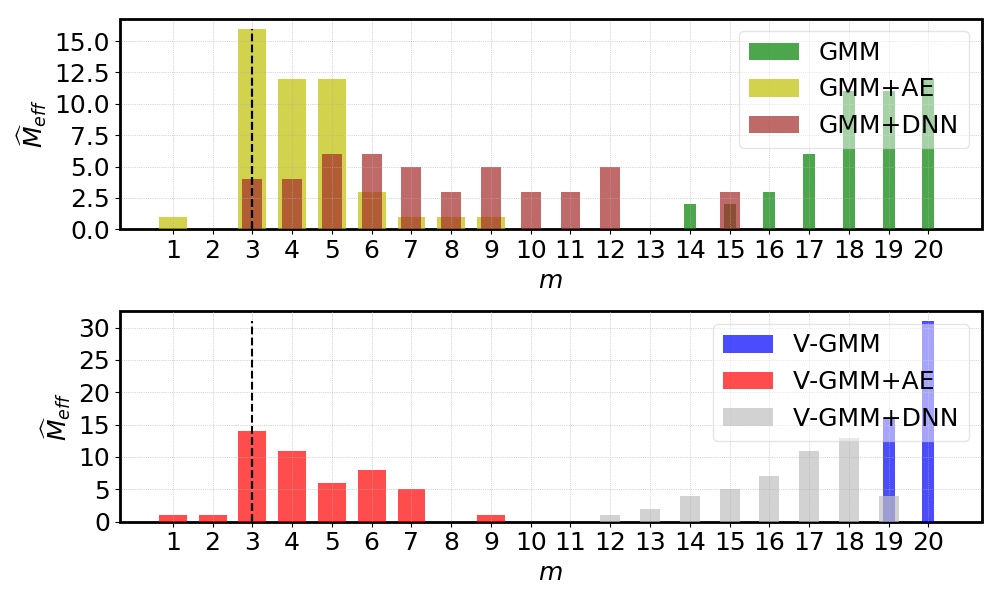}}
	\caption{
	 MC experiment given $N=15$. Histograms of $\widehat{M}$ (before
    rounding process) \textbf{(a)} and $\widehat{M}_{eff}$ (after
    rounding process) \textbf{(b)} over the $50$ MC runs - ground truth is identified by the vertical dashed line. 
	}
	\label{fig:reconstruction_MC}
\end{figure*}

In addition, Figure~\ref{fig:weights_MC} confirms the overfitting tendency of all the considered SM identification methods, especially in correspondence to the variational approaches, for which the histograms are quite spread far from the optimal value (i.e., $ M=3$). 
From Figure~\ref{fig:weights_eff_MC}, instead, one can perceive the main differences between the three classes of solutions and this is due to the rounding operation (whose benefits have already been mentioned in Section~\ref{subsec:run}). In particular, by interpreting an histogram as an empirical probabilistic measure, it results that the designed  \mbox{(V-)GMM+AE} strategy guarantees the higher probability of having $\widehat{M}_{eff} \approx M$. 
% More specifically, 
It is also worth noting that the non-linear DNN-based lossless transformation improves the performance of the \mbox{(V-)GMM} methods; however, to further improve the results, it is necessary to cope with the emerging curse of dimensionality via an AE-based dimensionality reduction. 

\begin{figure*}[t!]
	\centering
	\subfloat[\label{fig:weights_MC_3}]{
		\includegraphics[width=0.49\textwidth]{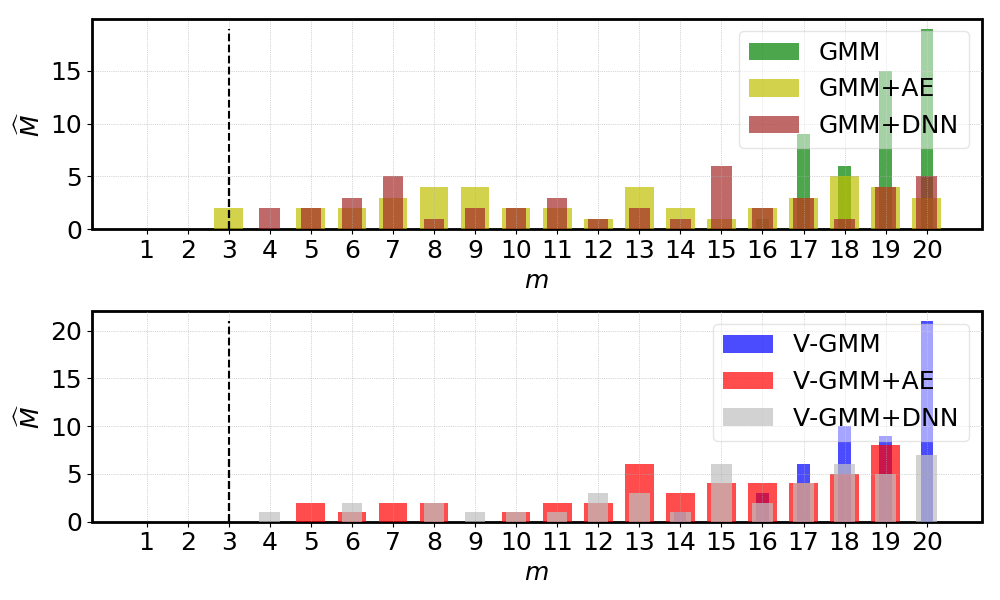}}%
	~ 
	\subfloat[\label{fig:weights_eff_MC_3}]{
		\includegraphics[width=0.49\textwidth]{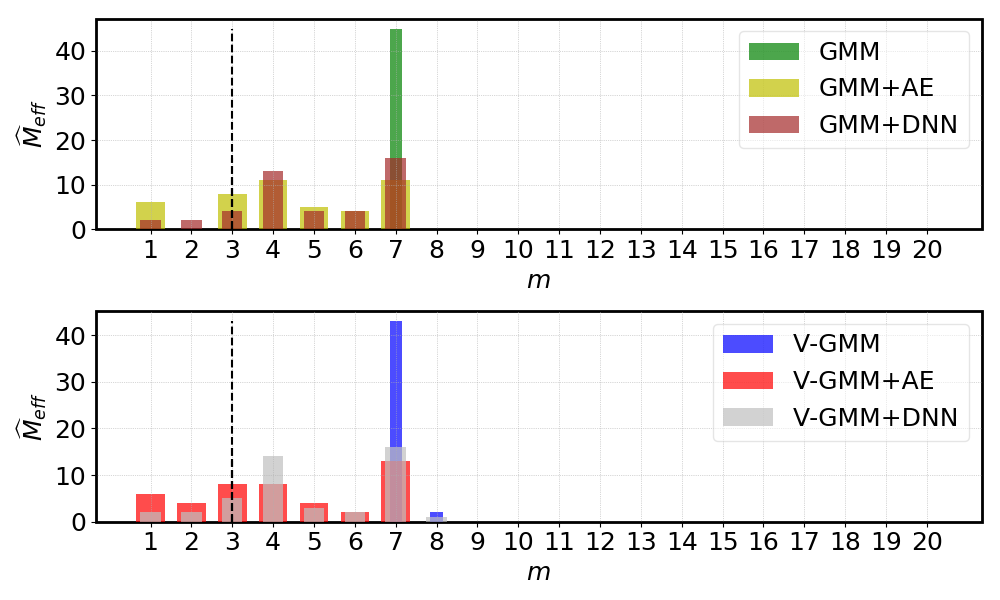}}
	\caption{ MC experiment given $N=3$. Histograms of $\widehat{M}$ (before
    rounding process) \textbf{(a)} and $\widehat{M}_{eff}$ (after
    rounding process) \textbf{(b)} over the $50$ MC runs - ground truth is identified by the vertical dashed line.
	}
	\label{fig:reconstruction_MC_3}
\end{figure*}

% %%%%%%%%%%%%%%%%%%%%%%%%%%%%%%%%%%%%%%%%%%%%%%%%
% \todo[inline]{------------}
% \todo[inline]{Questa discussione in caso si può eliminare, se non convince... noi puntiamo a large scale scenarios... oppure in alternativa teniamo solo la prima figura con gli istogrammi e non le ECDF... terrei invece il commento finale che comunque ci sta}

Finally, it is interesting to compare Figure~\ref{fig:reconstruction_MC} with Figure~\ref{fig:reconstruction_MC_3}. The latter reports the results gathered from a $50$-run MC experiment  accounting for a small VSN made up of $N=3$ cameras, given the parameters setup of Table~\ref{tab:setup_parameters}. 
One can realize that, for large scale networks (Figure~\ref{fig:reconstruction_MC}), the performance difference between the three considered classes of identification methods can be explained by the presence of the AE fulfilling a (beneficial) dimensionality reduction. 
In Figure~\ref{fig:reconstruction_MC_3}, indeed, the performance gap results to be less evident, intuitively because the dimensionality issues are milder, requiring also an encoder with simpler structure ($N-3-2$ layers, $p=2$). In detail, the \mbox{(V-)GMM+AE} and \mbox{(V-)GMM+DNN} methods turn out to behave similarly; on the other side, the (V-)GMM approach still experiences the overfitting phenomenon and its capability of reconstructing the events prior probability remains limited.  

In conclusion, it results that the proposed  (V-)GMM+AE  approach is advantageous if compared with the lossless (V-)\\GMM+DNN methods for the large scale VSNs of interest, because the drawbacks of the lossy AE transformation are compensated by the benefits of the dimensionality reduction. In addition, they anyway provide better performance with respect to the SM identification applied directly on the input data, namely performed through the (V-)GMM procedures, since in the latter case the beneficial effects of optimization or inference effect would not be present.

%%%%%%%%%%%%%%%%%%%%%%%%%%%%%%%%%%

% % EXPERIMENTS
% \section{Experiments on real datasets}\label{sec:experiments}
% \LV{Should we perform any experiment?}

%%%%%%%%%%%%%%%%%%%%%%%%%%%%%%%%%%

% CONCLUSION
\section{Conclusion}\label{sec:conclusion}
In particular, the suggested solution leverages a GMM approximation and an AE-based dimensionality reduction of the GMM learning problem. 
The algorithmic framework is discussed and theoretically presented.
Moreover, the identification procedure is first assessed with a simple numerical example and then it has been validated through more complete MC experiments on synthetic datasets.

The main innovative contribution with respect to the state of the art is the design of a model identification solution through an AE-based dimensionality reduction and GMM inference on the reduced feature space, which avoids the curse of dimensionality. The proposed procedure performance appears to be interesting compared to both a basic approach, where the model is obtained directly from the input data, and with respect to a lossless solution where a deep neural network is used to retain the original observation dimensionality. 

Future directions will be devoted to the extension of the proposed approach to distributed and dynamic scenarios (e.g., pan-tilt or mobile cameras); in particular, incremental GMM approaches may be used to update and refine the SM in time (\cite{zheng2019recursive}), %bigdeli2015incremental,
while concepts of federated machine learning may be applied to distribute the computation over the network~\cite{yang2019federated}. In addition, it is important to formally study the impact of the main setup parameters on the overall identification performance, as suggested by the results in Section~\ref{subsec:run}. Finally, it would be interesting to implement the proposed strategy on a real-world scenario and with some modifications on the overall pipeline described in Section~\ref{subsec:pipeline} (e.g., by adopting a Variational AE for the data encoding and dimensionality reduction process). 

%%%%%%%%%%%%%%%%%%%%%%%%%%%%%%%%%%

% REFERENCES
\bibliographystyle{IEEEtran}
\bibliography{IEEEabrv,References}

% Generated by IEEEtran.bst, version: 1.14 (2015/08/26)
\begin{thebibliography}{10}
\providecommand{\url}[1]{#1}
\csname url@samestyle\endcsname
\providecommand{\newblock}{\relax}
\providecommand{\bibinfo}[2]{#2}
\providecommand{\BIBentrySTDinterwordspacing}{\spaceskip=0pt\relax}
\providecommand{\BIBentryALTinterwordstretchfactor}{4}
\providecommand{\BIBentryALTinterwordspacing}{\spaceskip=\fontdimen2\font plus
\BIBentryALTinterwordstretchfactor\fontdimen3\font minus
  \fontdimen4\font\relax}
\providecommand{\BIBforeignlanguage}[2]{{%
\expandafter\ifx\csname l@#1\endcsname\relax
\typeout{** WARNING: IEEEtran.bst: No hyphenation pattern has been}%
\typeout{** loaded for the language `#1'. Using the pattern for}%
\typeout{** the default language instead.}%
\else
\language=\csname l@#1\endcsname
\fi
#2}}
\providecommand{\BIBdecl}{\relax}
\BIBdecl

\bibitem{kyung2016theory}
C.-M. Kyung \emph{et~al.}, \emph{Theory and applications of smart
  cameras}.\hskip 1em plus 0.5em minus 0.4em\relax Springer, 2016.

\bibitem{singh2021crowd}
U.~Singh, J.-F. Determe, F.~Horlin, and P.~De~Doncker, ``Crowd monitoring:
  State-of-the-art and future directions,'' \emph{IETE Technical Review},
  vol.~38, no.~6, pp. 578--594, 2021.

\bibitem{lissandrini2019cooperative}
N.~Lissandrini, G.~Michieletto, R.~Antonello, M.~Galvan, A.~Franco, and
  A.~Cenedese, ``Cooperative optimization of {UAV}s formation visual
  tracking,'' \emph{Robotics}, vol.~8, no.~3, p.~52, 2019.

\bibitem{liu2017robust}
L.~Liu, H.~Li, Y.~Dai, and Q.~Pan, ``Robust and efficient relative pose with a
  multi-camera system for autonomous driving in highly dynamic environments,''
  \emph{IEEE Transactions on Intelligent Transportation Systems}, vol.~19,
  no.~8, pp. 2432--2444, 2017.

\bibitem{mavrinac2013modeling}
A.~Mavrinac and X.~Chen, ``Modeling coverage in camera networks: A survey,''
  \emph{International journal of computer vision}, vol. 101, no.~1, pp.
  205--226, 2013.

\bibitem{esterle2014socio}
L.~Esterle, P.~R. Lewis, X.~Yao, and B.~Rinner, ``Socio-economic vision graph
  generation and handover in distributed smart camera networks,'' \emph{ACM
  Transactions on Sensor Networks (TOSN)}, vol.~10, no.~2, pp. 1--24, 2014.

\bibitem{dieber2011resource}
B.~Dieber, C.~Micheloni, and B.~Rinner, ``Resource-aware coverage and task
  assignment in visual sensor networks,'' \emph{IEEE Transactions on Circuits
  and Systems for Video Technology}, vol.~21, no.~10, pp. 1424--1437, 2011.

\bibitem{sanmiguel2014self}
J.~C. SanMiguel, C.~Micheloni, K.~Shoop, G.~L. Foresti, and A.~Cavallaro,
  ``Self-reconfigurable smart camera networks,'' \emph{Computer}, vol.~47,
  no.~5, pp. 67--73, 2014.

\bibitem{VarottoECC19}
L.~{Varotto}, M.~{Fabris}, G.~{Michieletto}, and A.~{Cenedese}, ``Distributed
  dual quaternion based localization of visual sensor networks,'' in \emph{2019
  18th European Control Conference (ECC)}, 2019, pp. 1836--1841.

\bibitem{sorrentino2020group}
F.~Sorrentino, L.~M. Pecora, and L.~Trajkovic, ``Group consensus in multilayer
  networks,'' \emph{IEEE Transactions on Network Science and Engineering},
  2020.

\bibitem{tesfaye2019multi}
Y.~T. Tesfaye, E.~Zemene, A.~Prati, M.~Pelillo, and M.~Shah, ``Multi-target
  tracking in multiple non-overlapping cameras using fast-constrained dominant
  sets,'' \emph{International Journal of Computer Vision}, vol. 127, no.~9, pp.
  1303--1320, 2019.

\bibitem{yang2019federated}
Q.~Yang, Y.~Liu, T.~Chen, and Y.~Tong, ``Federated machine learning: Concept
  and applications,'' \emph{ACM Transactions on Intelligent Systems and
  Technology (TIST)}, vol.~10, no.~2, pp. 1--19, 2019.

\bibitem{tzikas2008variational}
D.~G. Tzikas, A.~C. Likas, and N.~P. Galatsanos, ``The variational
  approximation for bayesian inference,'' \emph{IEEE Signal Processing
  Magazine}, vol.~25, no.~6, pp. 131--146, 2008.

\bibitem{GMM_search_static}
P.~Yao, Z.~Xie, and P.~Ren, ``Optimal uav route planning for coverage search of
  stationary target in river,'' \emph{IEEE Transactions on Control Systems
  Technology}, vol.~27, no.~2, pp. 822--829, 2017.

\bibitem{GMM_search_RHC}
P.~Yao, H.~Wang, and H.~Ji, ``Gaussian mixture model and receding horizon
  control for multiple uav search in complex environment,'' \emph{Nonlinear
  Dynamics}, vol.~88, no.~2, pp. 903--919, 2017.

\bibitem{GMM_tracking}
J.~M. Clark, P.-A. Kountouriotis, and R.~B. Vinter, ``A gaussian mixture filter
  for range-only tracking,'' \emph{IEEE transactions on automatic control},
  vol.~56, no.~3, pp. 602--613, 2010.

\bibitem{yan2021gmm}
S.~Yan and Y.~Dong, ``Gmm based simultaneous reconstruction and segmentation in
  x-ray ct application.'' in \emph{SSVM}.\hskip 1em plus 0.5em minus
  0.4em\relax Springer, 2021, pp. 503--515.

\bibitem{GMM_BS}
K.~Goyal and J.~Singhai, ``Review of background subtraction methods using
  gaussian mixture model for video surveillance systems,'' \emph{Artificial
  Intelligence Review}, vol.~50, no.~2, pp. 241--259, 2018.

\bibitem{GMM_speech}
D.~Povey, L.~Burget, M.~Agarwal, P.~Akyazi, K.~Feng, A.~Ghoshal, O.~Glembek,
  N.~K. Goel, M.~Karafi{\'a}t, A.~Rastrow \emph{et~al.}, ``Subspace gaussian
  mixture models for speech recognition,'' in \emph{2010 IEEE International
  Conference on Acoustics, Speech and Signal Processing}.\hskip 1em plus 0.5em
  minus 0.4em\relax IEEE, 2010, pp. 4330--4333.

\bibitem{zou2007determining}
X.~Zou, B.~Bhanu, B.~Song, and A.~K. Roy-Chowdhury, ``Determining topology in a
  distributed camera network,'' in \emph{2007 IEEE International Conference on
  Image Processing}, vol.~5.\hskip 1em plus 0.5em minus 0.4em\relax IEEE, 2007,
  pp. V--133.

\bibitem{Hussain2021multiview}
T.~Hussain, K.~Muhammad, W.~Ding, J.~Lloret, S.~W. Baik, and V.~H.~C. {de
  Albuquerque}, ``A comprehensive survey of multi-view video summarization,''
  \emph{Pattern Recognition}, vol. 109, p. 107567, 2021.

\bibitem{SpagnoloAghajanBebis2021}
P.~Spagnolo, H.~Aghajan, G.~Bebis, S.~Gong, A.~Loutfi, L.~Sigal, and W.-S.
  Zheng, ``Guest editorial introduction to the special issue on large-scale
  visual sensor networks: Architectures and applications,'' \emph{IEEE
  Transactions on Circuits and Systems for Video Technology}, vol.~31, no.~4,
  pp. 1249--1252, 2021.

\bibitem{WongCicekSoatto2021}
A.~Wong, S.~Cicek, and S.~Soatto, ``Learning topology from synthetic data for
  unsupervised depth completion,'' \emph{IEEE Robotics and Automation Letters},
  vol.~6, no.~2, pp. 1495--1502, 2021.

\bibitem{liu2021multi}
Q.~Liu, R.~Tahir, L.~K. Eric, L.~He \emph{et~al.}, ``Multi-camera logical
  topology inference via conditional probability graph convolution network,''
  in \emph{2021 IEEE International Conference on Multimedia and Expo
  (ICME)}.\hskip 1em plus 0.5em minus 0.4em\relax IEEE, 2021, pp. 1--6.

\bibitem{altahir2017modeling}
A.~A. Altahir, V.~S. Asirvadam, N.~H. Hamid, P.~Sebastian, N.~Saad, R.~Ibrahim,
  and S.~C. Dass, ``Modeling multicamera coverage for placement optimization,''
  \emph{IEEE sensors letters}, vol.~1, no.~6, pp. 1--4, 2017.

\bibitem{kritter2019optimal}
J.~Kritter, M.~Br{\'e}villiers, J.~Lepagnot, and L.~Idoumghar, ``On the optimal
  placement of cameras for surveillance and the underlying set cover problem,''
  \emph{Applied Soft Computing}, vol.~74, pp. 133--153, 2019.

\bibitem{han2019camera}
Z.~Han, S.~Li, C.~Cui, H.~Song, Y.~Kong, and F.~Qin, ``Camera planning for area
  surveillance: A new method for coverage inference and optimization using
  location-based service data,'' \emph{Computers, Environment and Urban
  Systems}, vol.~78, p. 101396, 2019.

\bibitem{cheng2019data}
K.~Cheng, M.~S. Khokhar, Q.~Liu, R.~Tahir, and M.~Li, ``Data-driven logical
  topology inference for managing safety and re-identification of patients
  through multi-cameras iot,'' \emph{IEEE Access}, vol.~7, pp.
  159\,466--159\,478, 2019.

\bibitem{cho2019joint}
Y.-J. Cho, S.-A. Kim, J.-H. Park, K.~Lee, and K.-J. Yoon, ``Joint person
  re-identification and camera network topology inference in multiple
  cameras,'' \emph{Computer Vision and Image Understanding}, vol. 180, pp.
  34--46, 2019.

\bibitem{LuvisottoClustering}
A.~{Cenedese}, M.~{Luvisotto}, and G.~{Michieletto}, ``Distributed clustering
  strategies in industrial wireless sensor networks,'' \emph{IEEE Transactions
  on Industrial Informatics}, vol.~13, no.~1, pp. 228--237, 2017.

\bibitem{javed2018community}
M.~A. Javed, M.~S. Younis, S.~Latif, J.~Qadir, and A.~Baig, ``Community
  detection in networks: A multidisciplinary review,'' \emph{Journal of Network
  and Computer Applications}, vol. 108, pp. 87--111, 2018.

\bibitem{milanfar2017super}
P.~Milanfar, \emph{Super-resolution imaging}.\hskip 1em plus 0.5em minus
  0.4em\relax CRC press, 2017.

\bibitem{varotto2021active}
L.~Varotto, A.~Cenedese, and A.~Cavallaro, ``Active sensing for search and
  tracking: A review,'' \emph{arXiv preprint arXiv:2112.02381}, 2021.

\bibitem{james2013introduction}
G.~James, D.~Witten, T.~Hastie, and R.~Tibshirani, \emph{An introduction to
  statistical learning}.\hskip 1em plus 0.5em minus 0.4em\relax Springer, 2013,
  vol. 112.

\bibitem{cenedese2010graph}
A.~Cenedese, R.~Ghirardello, R.~Guiotto, F.~Paggiaro, and L.~Schenato, ``On the
  graph building problem in camera networks,'' \emph{IFAC Proceedings Volumes},
  vol.~43, no.~19, pp. 299--304, 2010.

\bibitem{lucchese2014hidden}
R.~Lucchese, A.~Cenedese, and R.~Carli, ``A hidden markov model based
  transitional description of camera networks,'' \emph{IFAC Proceedings
  Volumes}, vol.~47, no.~3, pp. 7394--7399, 2014.

\bibitem{farrell2007learning}
R.~Farrell, D.~Doermann, and L.~S. Davis, ``Learning higher-order transition
  models in medium-scale camera networks,'' in \emph{2007 IEEE 11th
  International Conference on Computer Vision}.\hskip 1em plus 0.5em minus
  0.4em\relax IEEE, 2007, pp. 1--8.

\bibitem{cheng2006determining}
Z.~Cheng, D.~Devarajan, and R.~J. Radke, ``Determining vision graphs for
  distributed camera networks using feature digests,'' \emph{EURASIP Journal on
  Advances in Signal Processing}, vol. 2007, no.~1, p. 057034, 2006.

\bibitem{huang2018incorporating}
K.~Huang, Z.~Wang, and M.~Jusup, ``Incorporating latent constraints to enhance
  inference of network structure,'' \emph{IEEE Transactions on Network Science
  and Engineering}, 2018.

\bibitem{shi2020recovering}
L.~Shi, C.~Shen, Q.~Shi, Z.~Wang, J.~Zhao, X.~Li, and S.~Boccaletti,
  ``Recovering network structures based on evolutionary game dynamics via
  secure dimensional reduction,'' \emph{IEEE Transactions on Network Science
  and Engineering}, 2020.

\bibitem{charte2018practical}
D.~Charte, F.~Charte, S.~Garc{\'\i}a, M.~J. del Jesus, and F.~Herrera, ``A
  practical tutorial on autoencoders for nonlinear feature fusion: Taxonomy,
  models, software and guidelines,'' \emph{Information Fusion}, vol.~44, pp.
  78--96, 2018.

\bibitem{lai2021video}
Z.~Lai, S.~Liu, A.~A. Efros, and X.~Wang, ``Video autoencoder: self-supervised
  disentanglement of static 3d structure and motion,'' in \emph{Proceedings of
  the IEEE/CVF International Conference on Computer Vision}, 2021, pp.
  9730--9740.

\bibitem{cheng2019visual}
X.~Cheng, Y.~Zhang, L.~Zhou, and Y.~Zheng, ``Visual tracking via auto-encoder
  pair correlation filter,'' \emph{IEEE Transactions on Industrial
  Electronics}, vol.~67, no.~4, pp. 3288--3297, 2019.

\bibitem{wang2016auto}
Y.~Wang, H.~Yao, and S.~Zhao, ``Auto-encoder based dimensionality reduction,''
  \emph{Neurocomputing}, vol. 184, pp. 232--242, 2016.

\bibitem{li2018clustering}
Y.~Li, J.~Zhang, Z.~Ma, and Y.~Zhang, ``Clustering analysis in the wireless
  propagation channel with a variational gaussian mixture model,'' \emph{IEEE
  Transactions on Big Data}, 2018.

\bibitem{DEC}
J.~Xie, R.~Girshick, and A.~Farhadi, ``Unsupervised deep embedding for
  clustering analysis,'' in \emph{International conference on machine
  learning}, 2016, pp. 478--487.

\bibitem{steinbach2004}
M.~Steinbach, L.~Ert{\"o}z, and V.~Kumar, ``The challenges of clustering high
  dimensional data,'' in \emph{New directions in statistical physics}.\hskip
  1em plus 0.5em minus 0.4em\relax Springer, 2004, pp. 273--309.

\bibitem{dimensionalityReduction_tutorial}
A.~Ghodsi, ``Dimensionality reduction a short tutorial,'' \emph{Department of
  Statistics and Actuarial Science, Univ. of Waterloo, Ontario, Canada},
  vol.~37, no.~38, p. 2006, 2006.

\bibitem{dimensionalityReduction_review}
L.~Van Der~Maaten, E.~Postma, and J.~Van~den Herik, ``Dimensionality reduction:
  a comparative,'' \emph{J Mach Learn Res}, vol.~10, no. 66-71, p.~13, 2009.

\bibitem{DL_tutorial}
Q.~V. Le \emph{et~al.}, ``A tutorial on deep learning part 2: Autoencoders,
  convolutional neural networks and recurrent neural networks,'' \emph{Google
  Brain}, pp. 1--20, 2015.

\bibitem{goodfellow2016deep}
I.~Goodfellow, Y.~Bengio, A.~Courville, and Y.~Bengio, \emph{Deep
  learning}.\hskip 1em plus 0.5em minus 0.4em\relax MIT press Cambridge, 2016,
  vol.~1.

\bibitem{ahmad2019deep}
K.~Ahmad and N.~Conci, ``How deep features have improved event recognition in
  multimedia: a survey,'' \emph{ACM Transactions on Multimedia Computing,
  Communications, and Applications (TOMM)}, vol.~15, no.~2, pp. 1--27, 2019.

\bibitem{jin2020identification}
D.~Jin, X.~Wang, M.~Liu, J.~Wei, W.~Lu, and F.~Fogelman-Soulie,
  ``Identification of generalized semantic communities in large social
  networks,'' \emph{IEEE Transactions on Network Science and Engineering},
  2020.

\bibitem{wang2019temporal}
W.~Wang and X.~Li, ``Temporal stable community in time-varying networks,''
  \emph{IEEE Transactions on Network Science and Engineering}, 2019.

\bibitem{yolo}
J.~Redmon, S.~Divvala, R.~Girshick, and A.~Farhadi, ``You only look once:
  Unified, real-time object detection,'' in \emph{Proceedings of the IEEE
  conference on computer vision and pattern recognition}, 2016, pp. 779--788.

\bibitem{yoon2019structural}
J.~H. Yoon, C.-R. Lee, M.-H. Yang, and K.-J. Yoon, ``Structural constraint data
  association for online multi-object tracking,'' \emph{International Journal
  of Computer Vision}, vol. 127, no.~1, pp. 1--21, 2019.

\bibitem{sanmiguel2017efficient}
J.~C. SanMiguel and A.~Cavallaro, ``Efficient estimation of target detection
  quality,'' in \emph{2017 IEEE International Conference on Image Processing
  (ICIP)}.\hskip 1em plus 0.5em minus 0.4em\relax IEEE, 2017, pp. 915--919.

\bibitem{krivanek2003fast}
J.~Kriv{\'a}nek, J.~Zara, and K.~Bouatouch, ``Fast depth of field rendering
  with surface splatting,'' in \emph{Proceedings Computer Graphics
  International 2003}.\hskip 1em plus 0.5em minus 0.4em\relax IEEE, 2003, pp.
  196--201.

\bibitem{distributedEM_finite}
B.~Safarinejadian, M.~B. Menhaj, and M.~Karrari, ``A distributed em algorithm
  to estimate the parameters of a finite mixture of components,''
  \emph{Knowledge and information systems}, vol.~23, no.~3, pp. 267--292, 2010.

\bibitem{rothenberg1971identification}
T.~J. Rothenberg, ``Identification in parametric models,'' \emph{Econometrica:
  Journal of the Econometric Society}, pp. 577--591, 1971.

\bibitem{VarottoSelection}
L.~{Varotto}, A.~{Zampieri}, and A.~{Cenedese}, ``Street sensors set selection
  through road network modeling and observability measures,'' in \emph{2019
  27th Mediterranean Conference on Control and Automation (MED)}, 2019, pp.
  392--397.

\bibitem{poggio2018theory}
T.~Poggio and Q.~Liao, ``Theory i: Deep networks and the curse of
  dimensionality,'' \emph{Bulletin of the Polish Academy of Sciences. Technical
  Sciences}, vol.~66, no.~6, 2018.

\bibitem{li2021softpartitionclustering}
K.~Li, T.~Ni, J.~Xue, and Y.~Jiang, ``Deep soft clustering: simultaneous deep
  embedding and soft-partition clustering,'' \emph{Journal of Ambient
  Intelligence and Humanized Computing}, pp. 1--13, 2021.

\bibitem{tian2014learning}
F.~Tian, B.~Gao, Q.~Cui, E.~Chen, and T.-Y. Liu, ``Learning deep
  representations for graph clustering.'' in \emph{Aaai}, vol.~14.\hskip 1em
  plus 0.5em minus 0.4em\relax Citeseer, 2014, pp. 1293--1299.

\bibitem{caron2018deep}
M.~Caron, P.~Bojanowski, A.~Joulin, and M.~Douze, ``Deep clustering for
  unsupervised learning of visual features,'' in \emph{Proceedings of the
  European Conference on Computer Vision (ECCV)}, 2018, pp. 132--149.

\bibitem{AE_keras}
F.~Chollet, ``Building autoencoders in keras,'' \emph{The Keras Blog}, 2016.

\bibitem{zheng2019recursive}
J.~Zheng, Q.~Wen, and Z.~Song, ``Recursive gaussian mixture models for adaptive
  process monitoring,'' \emph{Industrial \& Engineering Chemistry Research},
  vol.~58, no.~16, pp. 6551--6561, 2019.

\end{thebibliography}

\newpage

%% Loading bibliography style file
%\bibliographystyle{model1-num-names}
%\bibliographystyle{cas-model2-names}

% Loading bibliography database
%\bibliography{}

% Biography

\bio{}
\bio{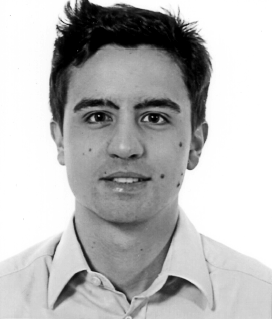}
\textbf{Luca Varotto} received the M.S. degree in automation engineering from the University of Padova, Italy, in 2018. From 2018 to 2021, he was a Ph.D. student at the Department of Information Engineering, University of Padova. In 2020, he has been a Visiting Researcher at the Centre for Intelligent Sensing, Queen Mary University of London, England. His research interests are related to active and multi-modal sensing, intelligent control for autonomous systems, and camera networks.
\endbio

\bio{}
\bio{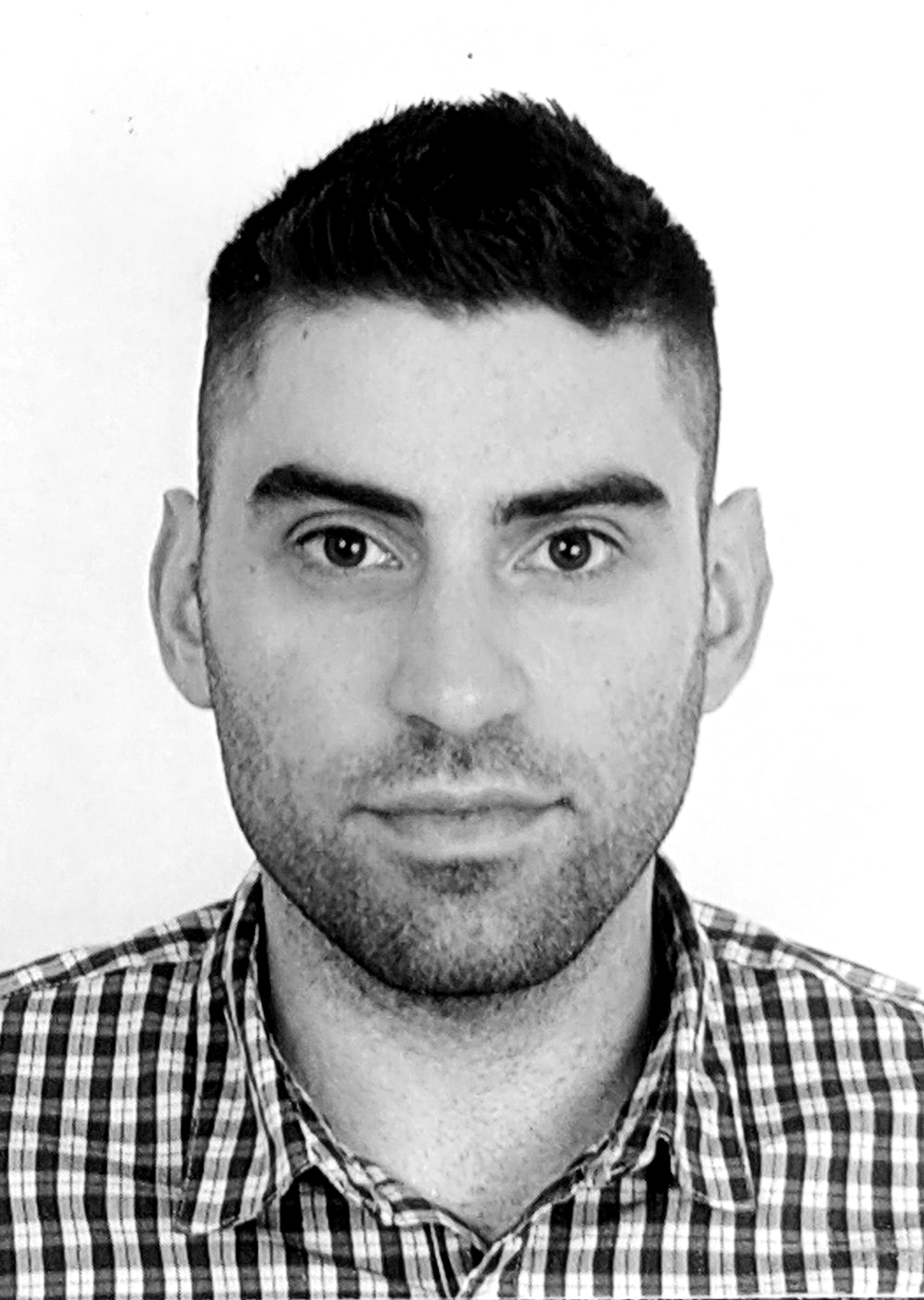}
\textbf{Marco Fabris} received the Laurea (M.Sc.) degree (summa cum laude) in Automation Engineering and his PhD both from the University of Padua, in 2016 and 2020, respectively. In 2018, he spent six months at the University of Colorado Boulder, USA, as a visiting scholar, focusing on distance-based formation control. In 2020-2021, he has been post-doctoral fellow at the Technion-Israel Institute of Technology, Haifa. His current research interests also involve graph-based consensus theory and optimal decentralized control and estimation for networked systems.
\endbio

\bio{}
\bio{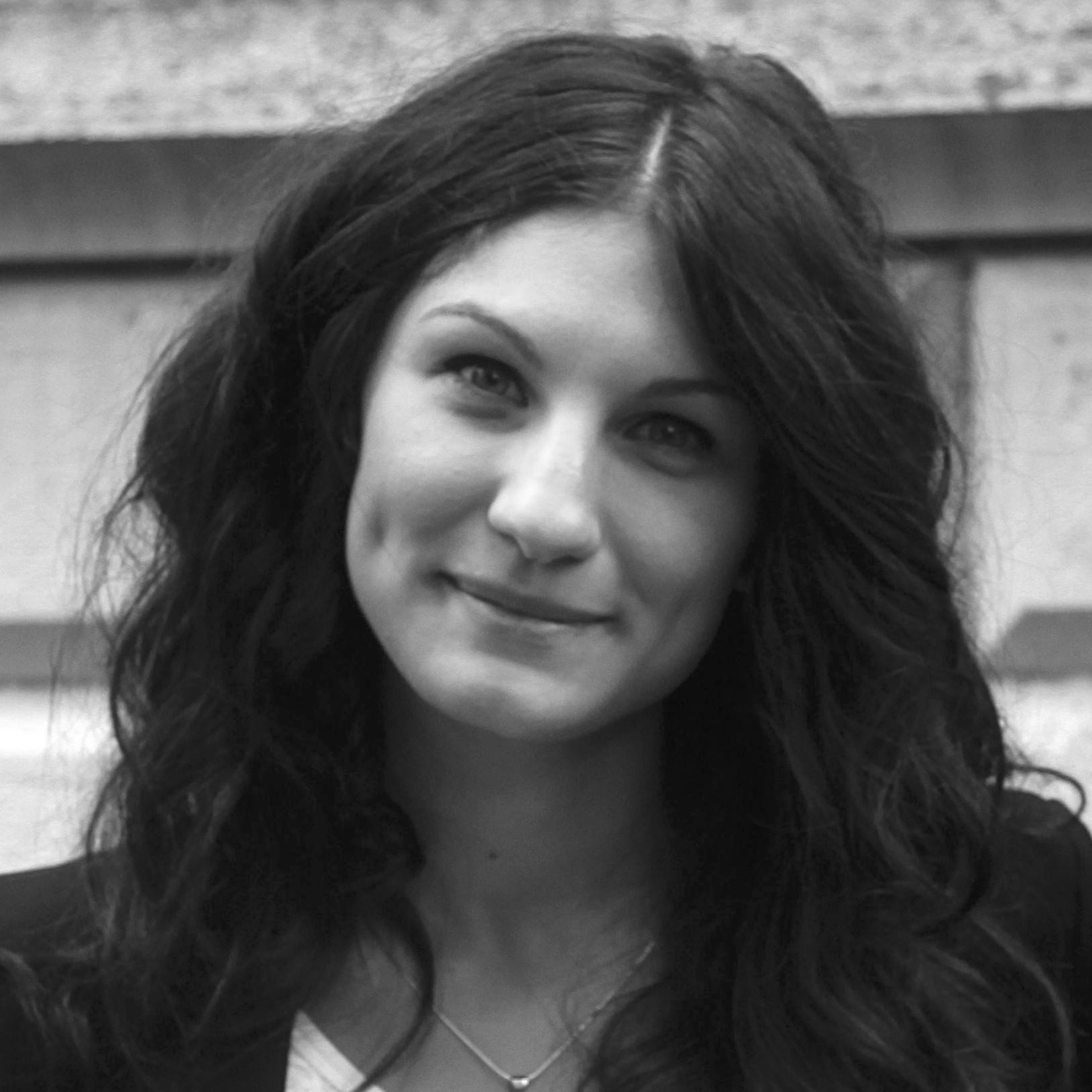}
\textbf{Giulia Michieletto} received the M.S. (2014) and the Ph.D. (2018) degrees from the University of Padova, Italy, from the University of Padova, Italy, where she is currently Assistant Professor with the Department of Management and Engineering. From March 2016 to February 2017, she was a Visiting Researcher at LAAS-CNRS, Toulouse, France.  From February 2018 to November 2019, she was a post-doc fellow with the SPARCS group at University of Padova, Italy.  Her main research interests include multi-agent systems modeling and control with a special regard to networked formations of ground and  aerial vehicles.
\endbio

\bio{}
\bio{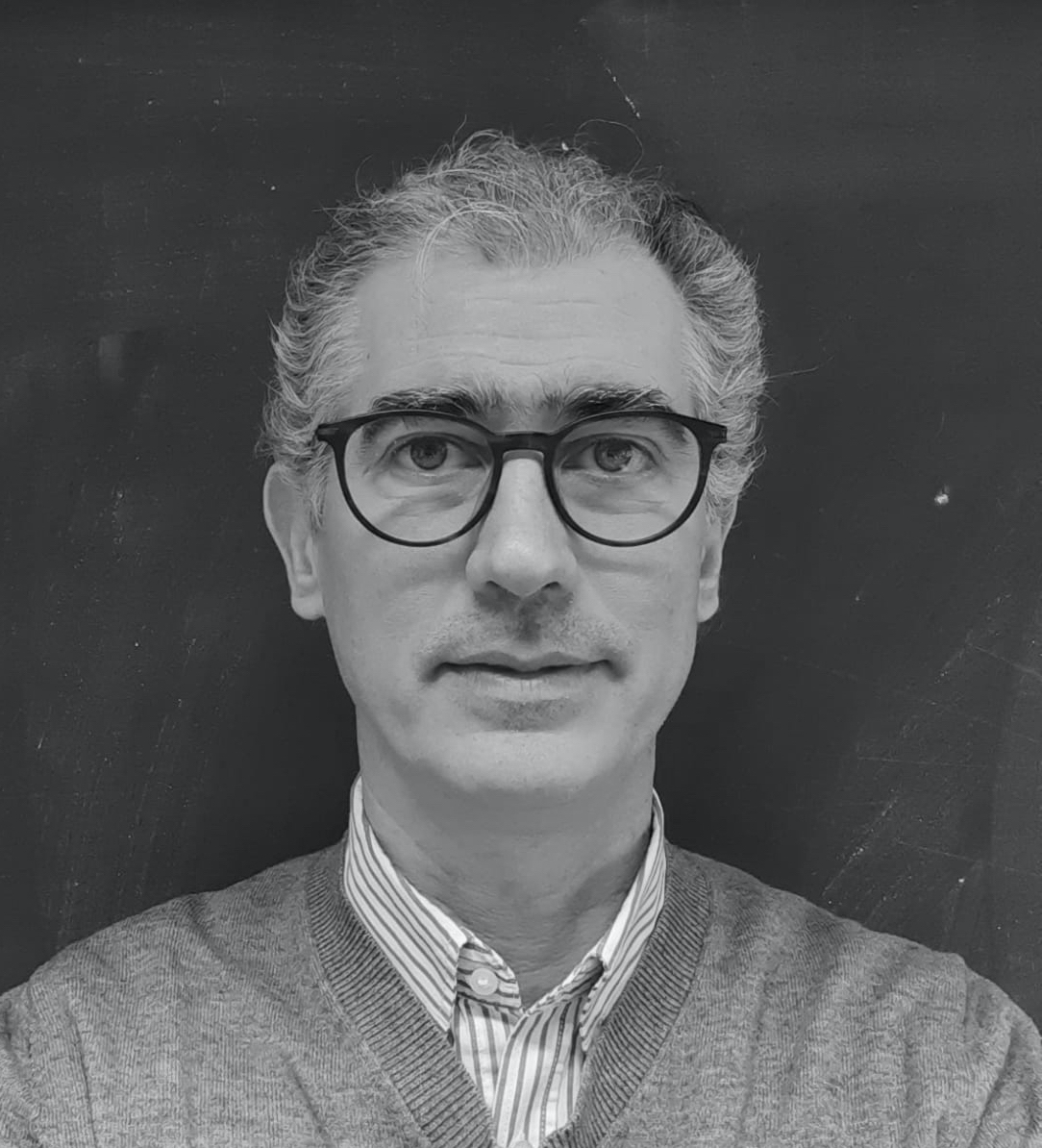}
\textbf{Angelo Cenedese} received the M.S. (1999) and the Ph.D. (2004) degrees from the University of Padova, Italy, where he is currently an Associate Professor with the Department of Information Engineering.
He is founder and leader of the research group SPARCS (SPace Aerial and gRound Control Systems) and responsible of the SPARCS Laboratory with flying arena and of the NAVLAB Industrial Application laboratory. He has been and he is involved in several projects on control of complex systems, funded by European and Italian government institutions and industries, with different roles of participant and/or principal investigator. His research interests include system modeling, control theory and its applications, mobile robotics, multi agent systems. On these subjects, he has published more than 180 papers and holds three patents.

\endbio

\end{document}